\documentstyle[prl,aps,epsf,multicol]{revtex}
\begin{document}
\draft
\preprint{TIFR : }
\title{ Pattern Formation in Interface Depinning and Other Models:
Erratically Moving Spatial Structures}
\author{Supriya Krishnamurthy and Mustansir Barma}
\address{Theoretical Physics Group, Tata Institute of Fundamental Research,
Homi Bhabha Road, Mumbai 400005, India}
\date{\today}
\maketitle
\begin{abstract}

We study erratically moving spatial structures that are found  in a driven interface
in a random medium at the depinning threshold.  We introduce a bond-disordered
variant of the Sneppen model and study the effect of extremal dynamics on
the morphology of the interface. We find evidence for the formation of 
a structure  which moves along with the growth site.  The time
average of the structure, which is defined with  respect to the active spot 
of growth, defines an activity-centered pattern.  Extensive Monte Carlo
simulations show that the pattern has a tail which decays slowly, as a power law.
To understand this sort of pattern formation, we write down an approximate
integral equation involving the local interface dynamics and long-ranged 
jumps of the growth spot.  
We clarify the nature of the approximation by considering a model for which the 
integral equation is exactly derivable from an extended master 
equation. Improvements to the equation are considered by adding a second
coupled equation which provides a self-consistent description. 
The pattern, which defines a one-point correlation function, 
is shown to have a strong effect on
ordinary space-fixed two-point correlation functions. 
Finally we present evidence that this sort of pattern formation is not
confined to the interface problem, but is generic to situations in which the
activity at succesive time steps is correlated, as for instance in several
other extremal models.  We present numerical results for activity-centered
patterns in the Bak-Sneppen model of evolution and the Zaitsev model 
of low-temperature creep.
\end{abstract}
\pacs{PACS numbers: 05.40.+j, 47.54.+r, 68.10.Gw, 47.55.Mh}

\begin{multicols}{2}

\section{Introduction}

Driven interfaces in  random media present several
features of interest, both with regard to the morphology of the moving
interface as well as the dynamics of the growth process. 
Experiments have been performed on several sorts of systems, ranging from
fluid flow in porous media \cite{skm,redg,hfv,hkw,buldyrevetal}
to propagation of burning fronts \cite{zhangetal}. These  indicate that the
disorder in the medium  
affects the properties of the interface in a crucial way. In particular,
the large-distance scaling properties differ considerably from those
of interfaces in uniform media. In both theoretical and experimental
investigations, it is customary to characterize the spatial structure of
the interface by its {\it roughness}. The main point of this paper
is to show that there is sometimes an unusual sort of pattern formation
\cite{skmb} in the system, which results in the interface acquiring a
{\it time-averaged shape}. In such situations, this pattern provides
an alternative characterization of interface morphology.

A customary measure of  the roughness is provided by the exponent
$\alpha$, defined by  $W \sim L^{\alpha}$,
where $W$ is the root mean squared width of the interface
and $L$ is the size of the system.  The experiments mentioned
above, and others similar to these, report an anomalously large value
of $\alpha$ --- large compared to the predictions of existing 
theories for interface growth in nonrandom media \cite {kpz}. 
It is recognised that the quenched nature of the 
disorder in the medium is responsible for this difference in
scaling properties of the interface.  Unlike thermal noise, which varies
rapidly, a portion of the interface subject to a quenched-noise environment
continues to experience the same forces until the growth process
takes it forward to a new region.  The pinning effect of quenched
noise has a strong effect on the large scale properties of the
interface. 

Several theoretical models have been put forward to try to account
for the effect of quenched disorder on the
properties of the interface; \cite{fisher} gives an account of some of the
early work and the relationship to other problems involving pinning,
while \cite{halzh} is a
a recent review. Amongst various proposals put forward to explain 
anomalous roughening are theories based on
continuum equations with quenched disorder
\cite{bruinsma,koplev,klt,parisi,natt,narfish}, 
the inclusion of noise with power-law
amplitude \cite{zhang,af} or long-ranged correlation
\cite{mhkz,meakinjullien,alf}, power-law distributions of
pinning-center strengths \cite{jensenprocaccia},
as well as a class of  
models with microscopic rules based on directed percolation
\cite{buldyrevetal,tang,sneppen} and models which relate the large-scale
structure to the wetting 
properties of the invading fluid \cite{martysetal}. 
A number of these models base the explanation of anomalous roughening
on the phenomenon of critical depinning, which is relevant to an
interface just at the threshold of motion. In the opposite limit of
large  velocity, the interface encounters
the disorder at any site only for a short time, suggesting that the quenched
nature of disorder is not important in this limit, and the
\vbox{
\epsfysize=7.0cm
\epsffile{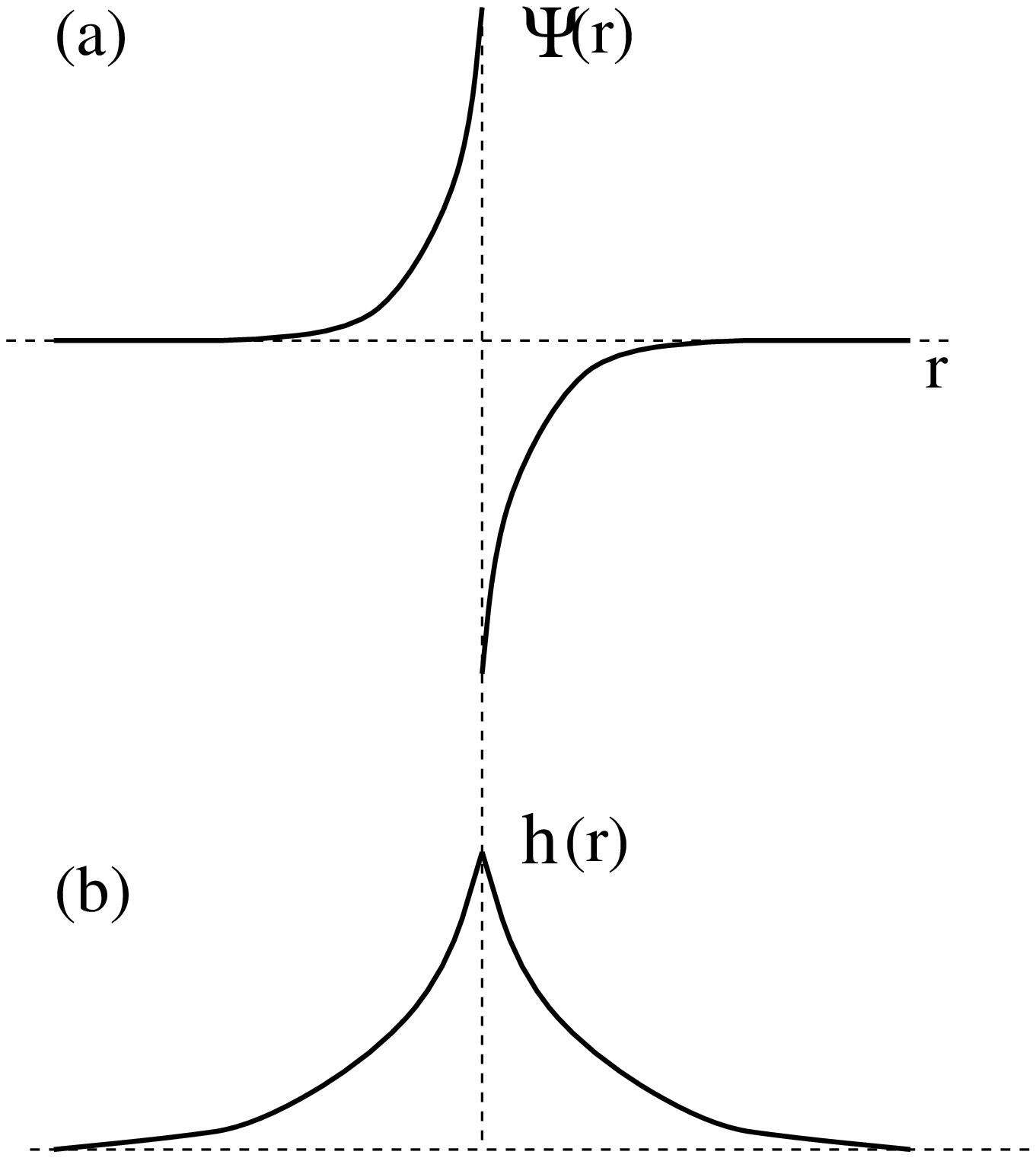}
\begin{figure}
\caption{\label{fig1}
\narrowtext
A schematic picture of the activity-centered pattern
in an {\em untilted} $(m =0)$ interface showing (a) the height gradient
pattern as defined in Eq. 1, (b) the interface profile
obtained by integrating the pattern in (a). 
The discontinuity at the origin in (a) leads to a cusp at the origin
in (b), implying that the active site is most likely at the peak.}
\end{figure}}
interface behaves much as in a nonrandom medium
\cite{parisi}. However this is not true at low velocities of the 
interface near the depinning threshold.
In particular, the limit of zero velocity is thought to be a
dynamical critical point
\cite{fisher} where the scaling properties of the
interface  are strongly affected by the disorder.

In a certain class of  models, the  quenched disorder enters as
barriers of random strengths which impede interface motion.
The formation of infinitely long  directed percolating paths of these
barriers 
is of special significance, as such paths can block the entire interface
effectively \cite{buldyrevetal,tang,sneppen,tang2}. 
The model proposed by Sneppen \cite{sneppen} involves ``extremal'' dynamics:
At each time step, the interface advances only along the weakest
of the barriers. Extremal models were first introduced much earlier,
in the context of invasion percolation for
two-phase fluid flow in porous media, with  a non-wetting fluid
displacing a wetting one \cite{lenormand,chandler,wilkinson}. 
The predictions of this invasion percolation model
were borne out by experiments
\cite{expts}.  Dynamical correlation functions involving the center of activity
in invasion percolation obey scaling \cite{furubergetal,rouxetal}, and
the process defines a
self-organised critical phenomenon \cite{rouxetal};
the interface organises itself to align along
critical paths bordered by large barriers, without
the necessity of tuning any external parameters.
The model proposed by Sneppen \cite{sneppen} is a modification of the
invasion percolation 
model, incorporating surface tension effects which prevent very strong
local convolutions of the interface, resulting in a self-affine, rather than
self-similar, geometry.
\vbox{
\epsfysize=7.0cm
\epsffile{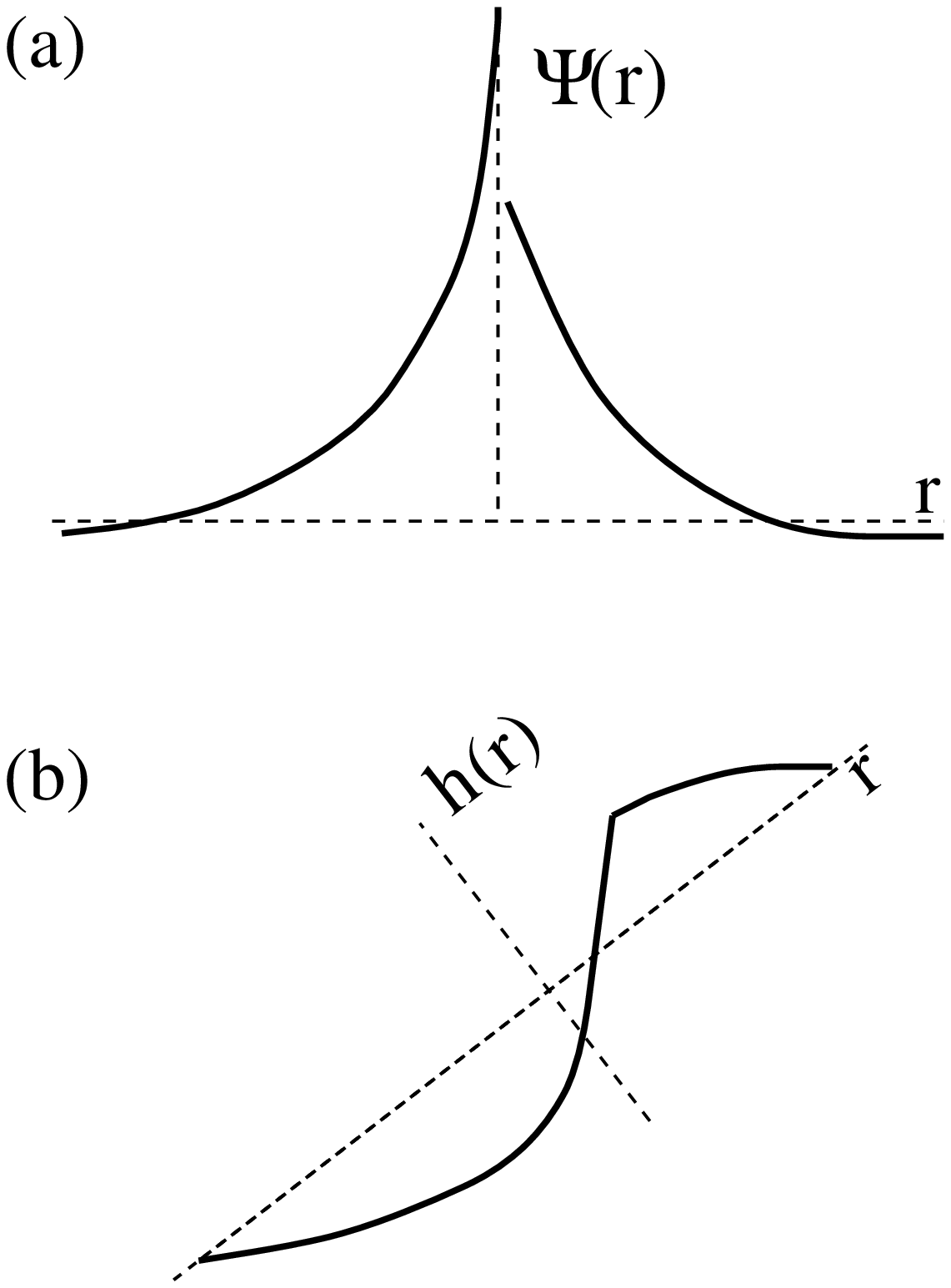}
\begin{figure}
\caption{\label{fig2}
\narrowtext
A schematic picture of the activity-centered pattern
in a {\em tilted} interface $(m \neq 0)$, in (a) height gradients
(b) the interface profile obtained by integrating the pattern
in (a).  As can be seen in
(b), there is a  larger than nominal slope near the active site.}
\end{figure}}
An extremal model close to the Sneppen model was shown to result 
\cite{roux-hansen}
from a model of wetting-fluid invasion, by considering an interface
advancing by merging  meniscus arcs between adjacent pairs of pinning 
centers.

An interesting feature of extremal dynamics is that it induces strong
spatial correlation in sites at which  growth occurs \cite{snjn,tl1,pmb} at successive
time steps. In this paper we introduce and study a variant of the
Sneppen model and show that the interface develops an interesting 
time-averaged structure as a result of correlations.
The defining equation for the structure is
\begin{equation}
\Psi (r) \equiv \frac{1}{2}[\langle \nabla h(r+R(t))\rangle - m]
\label{eq:Pattern} 
\end{equation} 
where $ R(t) $ is the position of the active site at time $t$,
$h(r^{\prime})$ denotes the height at the site $r^{\prime}$, $\langle
... \rangle $ is a time average in the steady state, and $m$ is the overall
slope of the interface. The unusual point is that
this structure  $\Psi (r)$
is not fixed in space, but moves with its center always at
the growth site, which itself follows an erratic path. The moving
origin is crucial to the definition, as time averages performed at
a fixed point in space reveal no structure at all. This average
structure defines  a pattern, and we study its
formation numerically and analytically. 

Figure 1a depicts schematically the activity-centered pattern in height
gradients with respect to the moving origin, for the untilted ($m=0$)
interface. As can be seen, the interface develops an
overall shape which is described by the height-gradient profile 
$\Psi (r)$ . The tail of the pattern falls slowly at large distances,
as a power law.
The corresponding height profile of the interface $h(r)$ with respect to the
active site is represented in Fig. 1b. The nature of the pattern is
sensitive to tilt, and 
Figure 2 shows the height-gradient pattern and height profile for a tilted 
interface.  

We propose that this pattern is a simple way of characterising a new
aspect of the morphology of the interface.  Traditional ways of
characterising the morphology involve, as has already been mentioned,
determining the roughness exponent $\alpha$. However such a definition
does not hone in on the overall {\it shape} of the interface.  In
situations such as the one considered in this paper when the interface
does develop a nontrivial structure, the pattern is a useful
quantitative characterization.  An important point about the
time-averaged pattern is that Eq. 1 defines a one-point correlation
function.  As such, it would be expected to strongly influence the
properties of customary two-point correlation functions. We verify this
by numerically studying two-point correlations. 
Further we find that this sort of pattern
formation is not restricted only to the Sneppen model, but also occurs
in other extremal models, such as 
the Zaitsev model for low-temperature creep \cite{zaitsev} and
the Bak-Sneppen model \cite{bak-sn} of biological evolution, albeit in other 
quantities.

The plan of this paper is as follows.
In Section II we introduce our model and discuss the 
connection with the problem of directed percolation, well
established from earlier studies. In Section III, we discuss the
correlations in the location of successive growth sites, a concept
central to this paper because of its connection with pattern
formation. In Sections IV and V we define the
pattern and present numerical results as well as an 
integral equation which provides an
understanding of this sort of pattern formation. We
define a model for which the equation is exact and discuss how the
approximation can be improved.
In Section VI, we discuss the issue of temporal correlations with a
view to seeing how they affect the pattern. In Section VII, we present our
results for an ordinary two-point correlation function in our model
and show that activity-centered pattern formation has to be taken into
account in 
order to understand some features in it. Section VIII deals with 
pattern formation in other extremal models and we conclude with a
summary of our results in Section IX.

\section { extremal model of interface depinning}

The extremal-model description of fluid-fluid interfaces in porous
media is valid when the wetting is dominated by capillary forces,
and thermal fluctuations are not important. 
In the extremal model, the interface advances
along the weakest barrier just ahead of it. The appealing feature of
the  model is that it is 
self-organised critical; the dynamics, which involves  
searches for the global minimum at every step, automatically tunes the
interface to a critical state at
the depinning transition, without the necessity of fixing any external
parameter.
\vbox{
\epsfysize=4.5cm
\epsffile{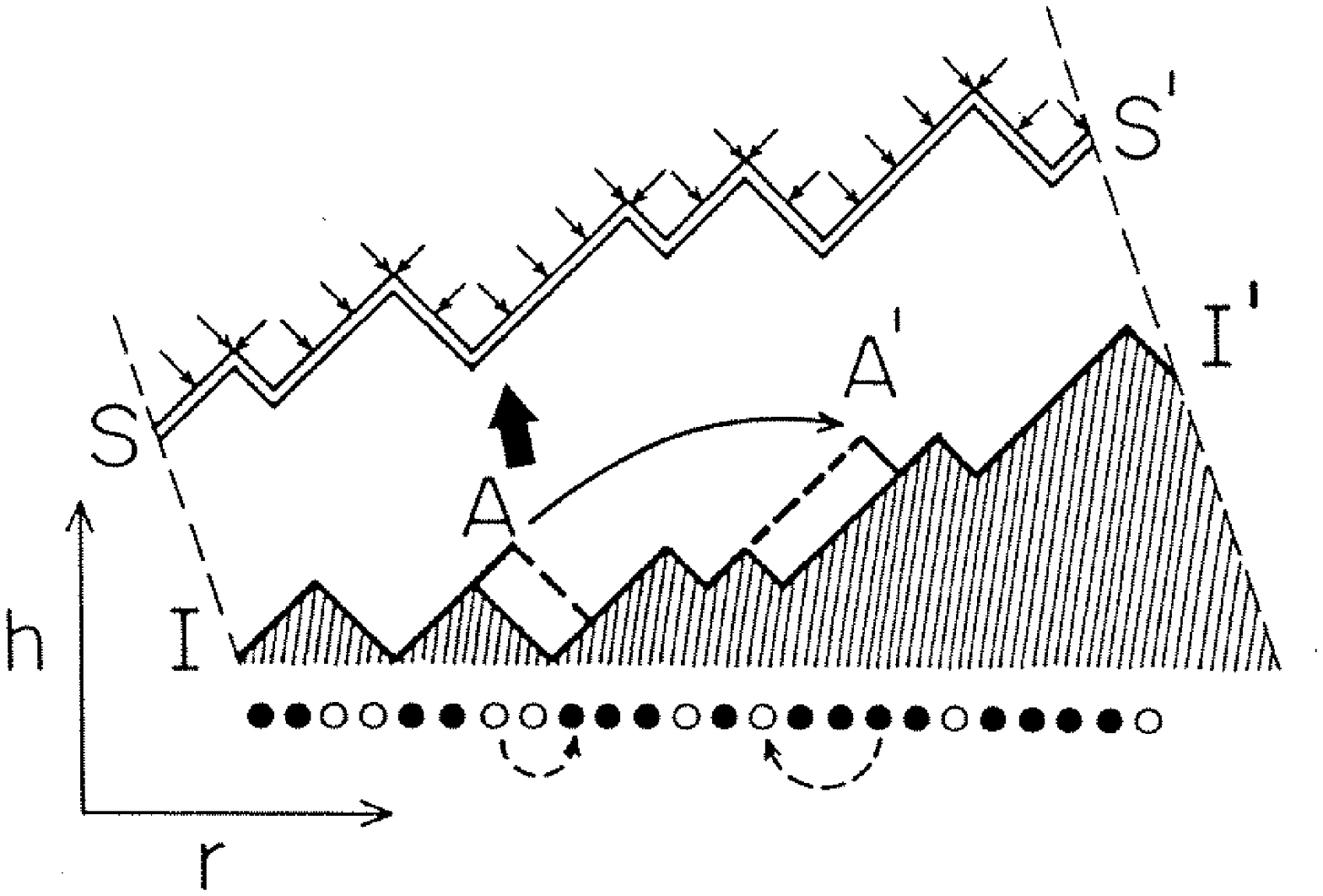}
\begin{figure}
\caption{\label{fig3}
\narrowtext
The Extremal Bond Model. The tilted  interface $II^{\prime}$ 
advances along the extremal perimeter
bond $A$ and locally readjusts to align along the dashed line. At the
next instant, the activity moves from $A$ to $A^{\prime}$. The corresponding
configuration and local moves for the particle-hole model are also
shown. $ SS^{\prime}$ is a stopper, whose perimeter is fully occupied by
diodes.}
\end{figure}}
In the model proposed by Sneppen \cite{sneppen},  
the  random medium is modelled by a square lattice in which the sites
are assigned random numbers $f \in [0,1]$. The random numbers could 
signify, for instance, the pore sizes in a porous medium. The
interface is a directed 
path on this lattice and grows only at that perimeter bond with the smallest
value of the random number; after every such move,  a local
rearrangement process \cite{kim-koster} ensures the absence of
very large slopes. 

Extremal
dynamics has also been proposed to describe very different
situations --- for instance, 
the phenomenon of low-temperature dislocation creep \cite
{zaitsev,roux-hansen}, crack propagation \cite{paczuski-peter} and
biological evolution \cite{bak-sn}.
We will see in Section VIII that the sort of pattern formation we find
in the interface model occurs in these models as well.

\subsection {The Extremal Bond Model}
We study a modified version of the Sneppen model
in this paper. In this version, hereafter referred to as
the Extremal Bond model (EBM), the interface is taken to be a directed
path on a 
square lattice (Fig. 3), with tilted cylindrical boundary
conditions \cite{mbjphysa} which ensure that the mean slope is
preserved. To every bond $k$ on the lattice is pre-assigned a fixed
random number $f_k$  drawn from the interval [0,1]. The interface
grows at that bond, in front of the interface, which carries the
smallest random number. The local growth rules are the following; if
the  chosen minimal bond has a positive (negative)
slope, the sequence of  links with negative (positive) slope just
below (on the left) also advances, as illustrated in Fig. 3.
This preserves the length of the interface as would
happen in situations with very high surface tension.
The local dynamics of interface adjustment is similar to that
for the low-noise Toom interface in ~\cite{dlss}; the models
are different in that the growth site is picked by the extremal
rule in our case, while it is picked stochastically in the Toom
interface model.
  
The EBM differs from the Sneppen model in that the length
of the interface is a strict constant of the motion and the $ f_{k}$'s
are associated with bonds rather than sites. While this modification
does not change the values of any of the large-distance scaling
properties of the
Sneppen model,  it has a few advantages. 
The interface aligns along directed spanning paths in a percolation
problem, just as in the Sneppen model. In our case, the
corresponding percolation problem is the diode-resistor percolation 
problem ~\cite{dhar-barma-phani}. 
On the square lattice, it is dual to the directed bond percolation 
problem ~\cite{kinzel} which is relatively well-studied.
Another advantage is that the  problem of interface growth in the
EBM is conceptually simplified by the existence of a known
one-to-one correspondence between the growing interface and a system
of hard-core 
particles moving on a ring. The two-dimensional problem hence
reduces to an effectively one-dimensional one. This also facilitates
the numerics. The correspondence between the interface and the
hard-core particles is detailed below.  

Positive slope links of the interface are represented by particles
($n_j =1$) and negative-slope links by holes ($n_j = 0$); see Fig. 3.
The difference in height of the interface between sites $j_{1}$ and
$j_{2}$ is given by $ h_{j_{2}} - h_{j_{1}} = \sum_{j=j_{1}}^{j_{2}}
(2 n_{j} - 1)$. 
In front of each link of the interface is a bond with
a random number $f$ assigned to it. Correspondingly, the site $j$ with
the particle (hole) representing this link carries a random number
$f_{j}$.  Just as for the interface, at each time step, activity is
initiated at the site with the minimum $f_{j}$. The update rules for
the interface translate to the following dynamics for particles and
holes. If the site with minimum $f_{j}$ contains a particle (hole), it
exchanges with the first hole (particle) to the left (right). All
sites hopped over, including the two which exchange the particle and
hole, are refreshed by assigning a new set of $f_{j}$'s. This
corresponds to the fact that the updated portion of the interface
moves ahead and meets a fresh set of $f$'s on the square
lattice. Because the number of positive slope links (and hence also
the number of negative slope links) is conserved for the interface, in
the particle-hole terminology, this implies that the number of
particles is conserved. Hence we can define a density $\rho $ for
particles on the one-dimensional lattice. This density determines the
mean slope $m=2\rho -1$ of the interface. An untilted interface
corresponds to half-filling. The reference direction for determining
tilt is the easy direction of directed bond percolation on the square
lattice, which is along the $45^{\circ}$ line. Tilt refers to any
density away from $0.5$, which implies a slope different from
$45^{\circ}$. The interface advances in a direction perpendicular to
the direction of tilt and this translates to a nonzero current of
particles on the ring.

\subsection{ Connection to Diode-Resistor Percolation and Directed
Percolation} 
Extremal models of interface depinning make use of 
a  correpondence to the
Diode Resistor Percolation (DRP) and Directed Percolation (DP) problems
to predict 
various properties of the interface. In view of this,  it is 
useful to recall some facts about the DP and DRP processes.
 
In the directed percolation problem, bonds on a 
lattice are occupied with probability $p$. At some critical value $p_c$
an infinite directed path of 
occupied bonds (in which every step is taken rightward or upward)
first forms along a definite direction; on a $2-d$ square lattice,
this is along the $45^{\circ}$ direction. For $p > p_c$, 
The network of these infinite paths forms an infinite connected cluster.
For directed bond percolation on a square lattice the value of $p_c$ is 
known to be $ \simeq 0.6446$ \cite{kinzel}. There are two 
distinct correlation lengths, 
$\xi_{\parallel}$ along the easy direction and $\xi_{\perp}$
transverse to it, both of which diverge as $p \rightarrow p_c$:
$\xi_{\parallel} \sim (p-p_c)^{- \nu_{\parallel}} ~, \xi_{\perp}
\sim (p-p_c)^{- \nu_{\perp}}$ respectively. The values of these
exponents are known  to be  $\nu_{\parallel} \simeq 1.733$ and
$\nu_{\perp} \simeq  1.097$ \cite{kinzel}. 

Suppose we have a single source point, and we ask which portion of
the plane can be reached from it via occupied directed bonds.
For $p>p_c$ this region is contained within a cone with
opening angle $2 \phi= 
\arctan (m)$ where $m$ is the slope of the edge of the cone
with respect to the  $45^{\circ}$ direction;  the 
opening angle depends on $p$. This relation can be inverted to find the
critical probability $p_c(m)$ {\it viz.} the probability at which a
connection first appears along the direction with slope $m\neq 0$.
Correlation lengths 
along and perpendicular to this direction have exponents
$\nu_{\parallel} = 1$ and $\nu_{\perp}= 0.5$ \cite{kinzel}.
We refer to the direction along $  45^{\circ}$ as
untilted ($m=0$);  any other slope is referred to as tilted. 

In the diode-resistor percolation problem, every bond is occupied 
by a ``diode'' (a one-way connection) with a probability $ p$  
or  a ``resistor'' (a two-way connection) with a probability $1-p$.  On
a square lattice, the diodes all point up or right.  Let us ask
which regions of the plane
are connected to a given source point. If $p=0$, a source point 
can reach the entire quadrant of which it is the left corner.
As $p$ decreases from $1$ to $p_c$, the opening angle $\phi ^{\prime}$ of
the connected region increases from $\pi /2$ to $\pi$; beyond this,
the entire plane can be reached from the source point. The edge of the
connected region is bordered 
by diodes pointing rightward and upward, which prevent it
from spreading leftward and downward (Fig. 4). 

On the $2-d$ square lattice, DP and DRP are dual to each other
\cite{dhar-barma-phani}. 
The
dual to a DRP configuration is constructed using the following rules.
A diode in the DRP lattice is
crossed by a diode in the dual lattice, whereas a resistor is crossed
by an insulator (no connection) in the dual lattice.
\vbox{
\epsfysize=7.4cm
\epsffile{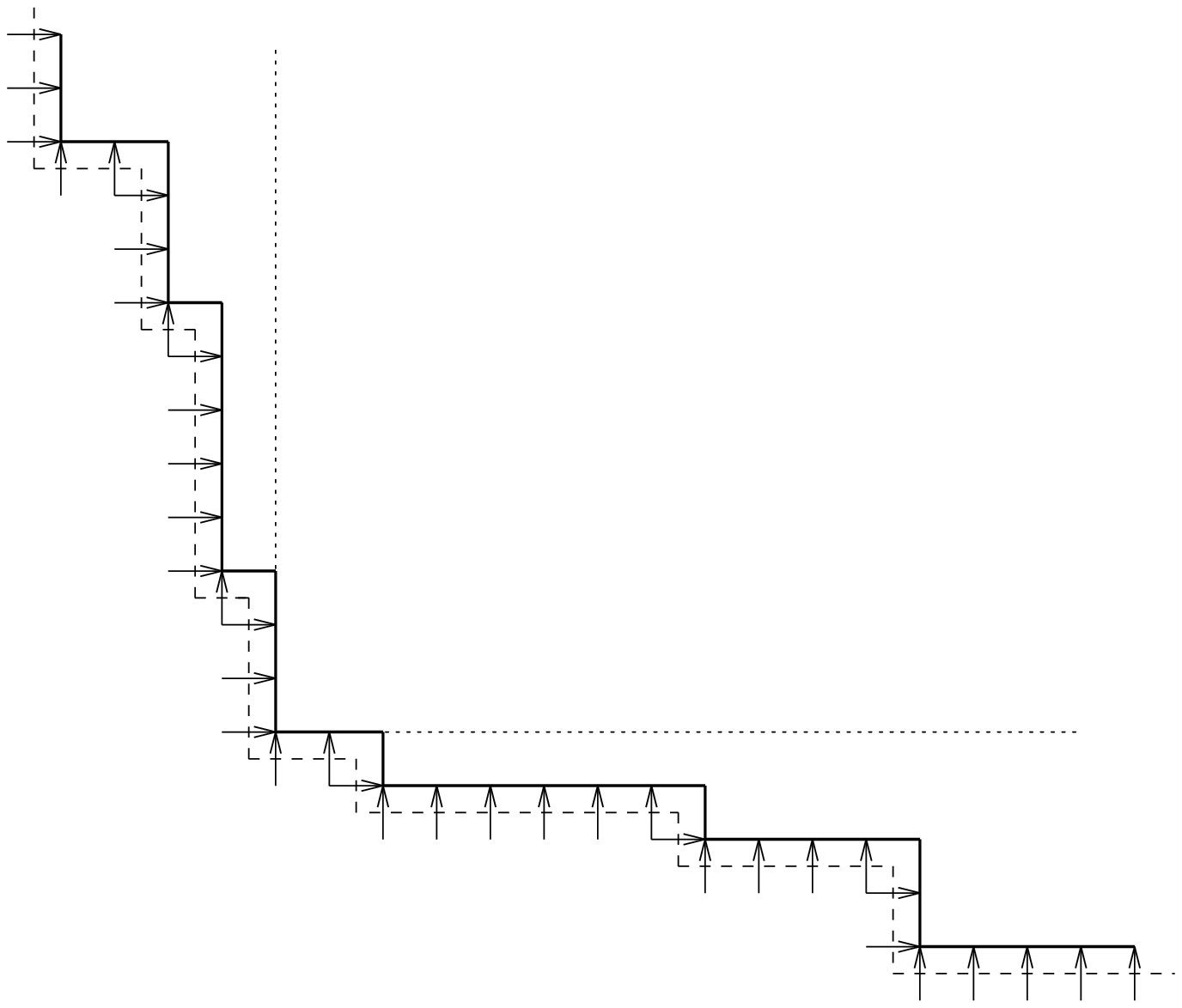}
\begin{figure}
\caption{\label{fig4}
\narrowtext
The boundary of the region reachable from the origin in 
diode-resistor percolation. For a nonzero fraction of resistors, the
opening angle of the region is larger than $90^{\circ}$ as shown. The
dashed line is an infinite path in the corresponding directed
percolation problem on the dual lattice.}
\end{figure}}
Thus we recover
the directed percolation problem on the dual lattice.  The opening
angle of the cones in the two problems are related by 
$ \phi + \phi ^{\prime} = {\pi}$.

 
In the EBM, the random
medium is modelled by considering a square lattice with every
bond assigned a random number $f_k$ drawn from the interval $[0,1]$. 
For a certain trial value $f^*$, imagine occupying all bonds with $f<f^*$
by resistors, and the rest with diodes. 
We thereby generate a DRP
configuration with $p=1-f^*$. When $f^*$ takes on the value $1-p_c$,
an infinite connected path of diodes is formed. Such a path is called
a `stopper', and is significant for the dynamics of the EBM, as a moving
interface with no overall tilt will align with such stoppers from time to
time \cite {tang2}. 
When the interface aligns along a stopper, all
the bonds in front of it are larger than $f_c=1-p_c$ (Fig. 5a).  
Similarly a tilted interface with slope $m$ aligns along the edge of the cone 
with the same slope and all the bonds in front of it are expected to 
have a value larger
than $f_c(m)=1-p_c(m)$. Since $p_c(m) > p_c(0)$ for $m \neq 0$, $f_c(m)
< f_c(0)$ (Fig. 5b). 
Even when the interface is evolving 
between two stoppers, only a small fraction of its overall length actually
is in between; the rest is still aligned with a stopper.  The non-aligned
fraction is expected to vanish in the thermodynamic limit.
These expectations are confirmed by numerical
studies of the EBM. As can be seen from the figures,  
the bonds in front of the interface are all mostly larger than a
threshold value. 

Consider an interface of slope $m$ aligned along a critical DRP
path of the same slope.
\vbox{
\epsfysize=5.5cm
\epsffile{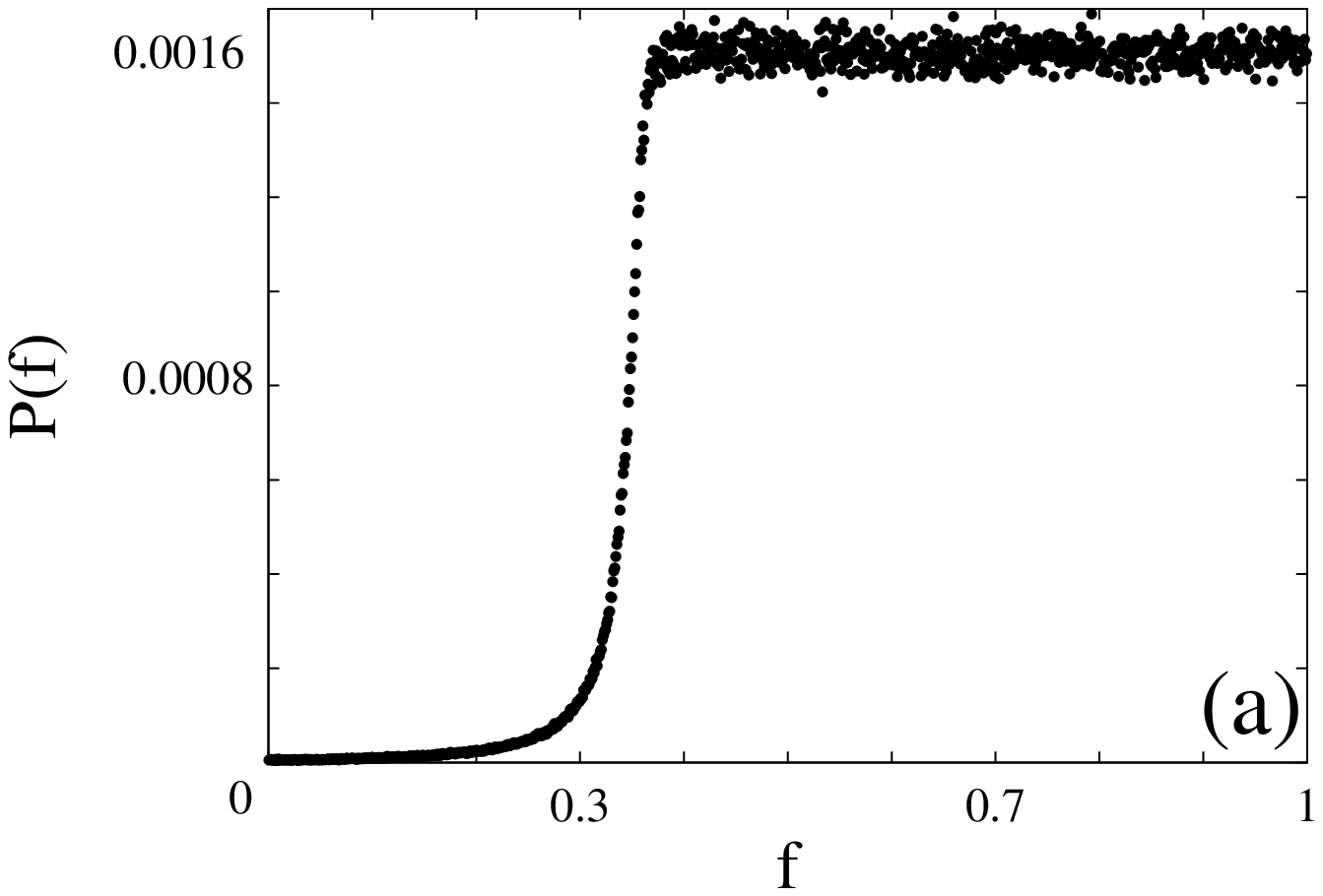}
\begin{figure}
\epsfysize=5.5cm
\epsffile{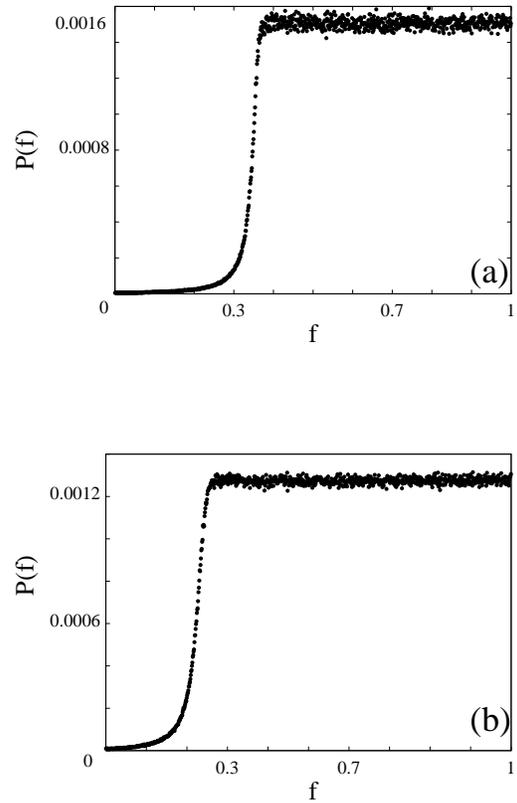}
\caption{\label{fig5}
\narrowtext
The steady state probability $P(f)$ of having a bond with
the value of the random number equal to $f$, in front of the
(a) untilted (b) tilted interface with $\rho =0.75$, in the Extremal
Bond Model. There 
is a tilt dependent threshold below which the probabilty of finding a
bond with that value is very low. A system of size $1000$ was 
averaged over $10^{7}$ configurations}
\end{figure}}
It then moves forward by puncturing the path at
the site with the least value of $f$, which for an infinite system 
is exactly $1-p_c(m)$. On piercing through,
a portion of the interface grows and fills out a loop of the
infinite cluster while the rest of it remains pinned. 
However the interface
motion within the loop is far from uniform. Just as the critical
cluster at $p_c(m)$ impedes the growth of the interface on length
scales of the order of the size of the system, near-critical clusters 
impede its motion at length scales of the order of but smaller than
the loop size. One can think
of these clusters as forming a finer network of connections within
the network formed by the critical cluster at $p_c(m)$. 
While the interface is filling out a loop of the critical
cluster, it encounters this finer mesh  and as a result its motion is
impeded temporarily. In what follows, we refer to these near-critical
connections as `sub-stoppers'. A substopper can be characterized by
the lowest value of $f_k$ on it, say $f^{ss}$, and
also by the typical length scale $l$ over which it provides for effective 
pinning of the interface. These are related through
$ | f^{ss}-f_c(m)|^{-\nu_{\parallel}} \sim l$. 

\vbox{
\epsfysize=5.2cm
\epsffile{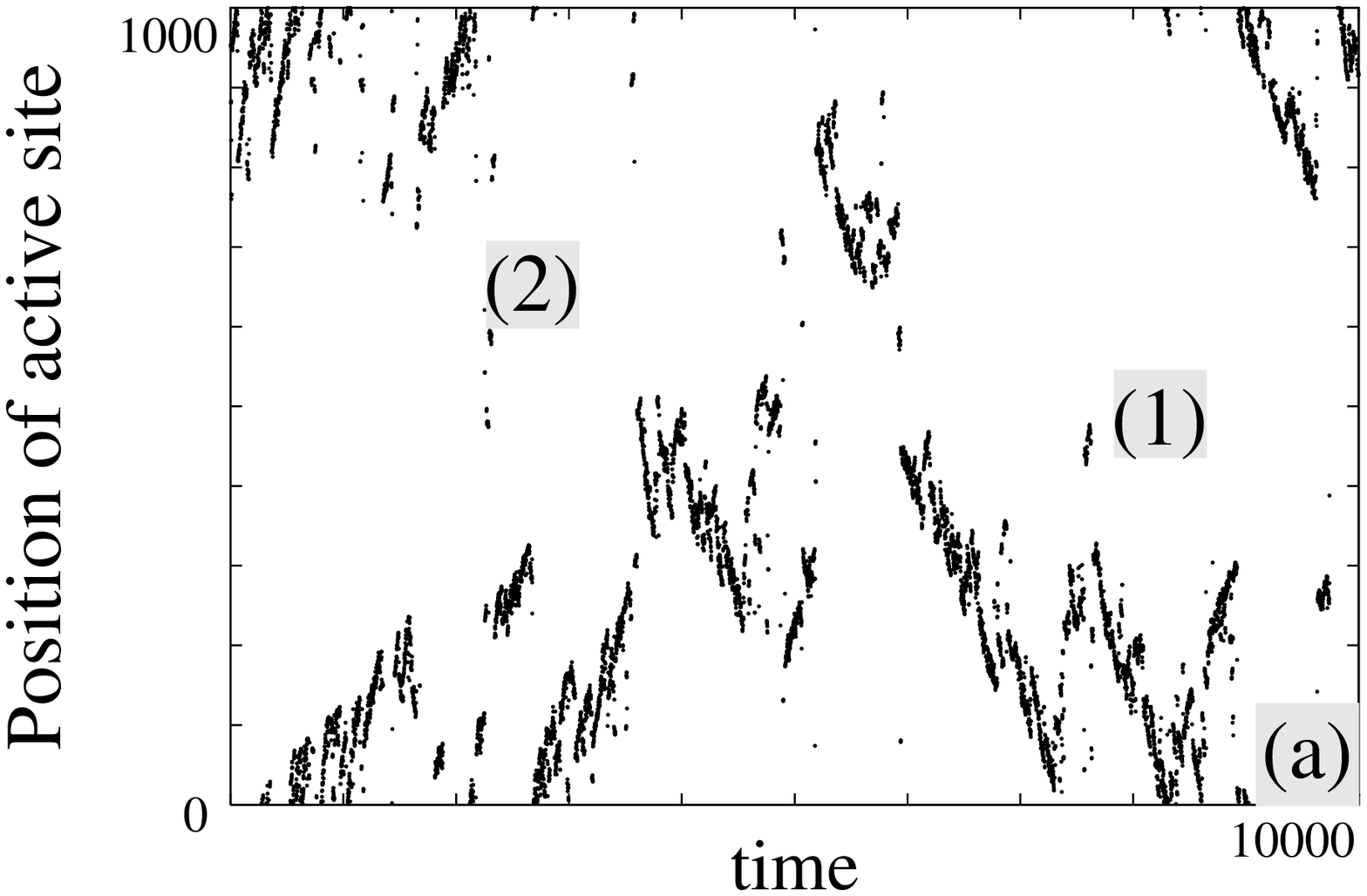}
\begin{figure}
\epsfysize=5.2cm
\epsffile{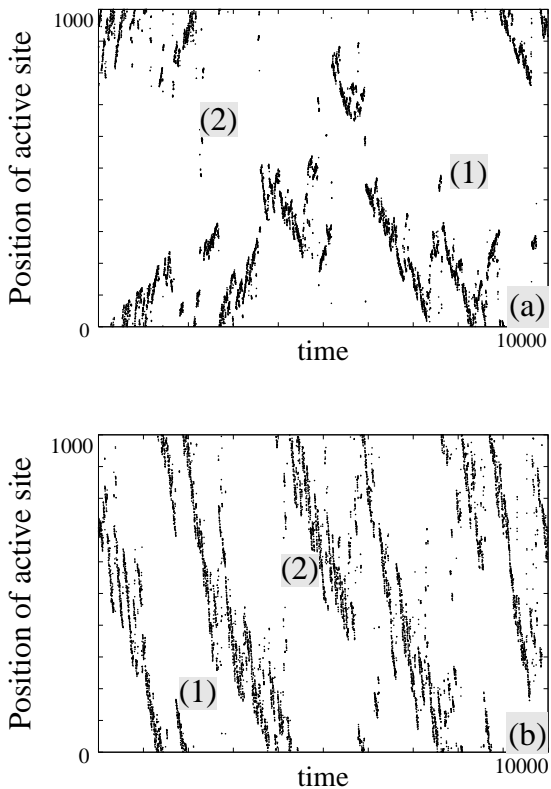}
\caption{\label{fig6}
\narrowtext
The location of the active site as a function of time for 
(a) untilted (b) tilted interface. The region marked (1) is a typical
example of a window in time in which the activity is localised in a
region. In (2) the activity ranges through the whole system,
corresponding to an instance when the interface has been pinned by a
critical cluster at $p_{c}(m)$. Though the directionality of the
active-site motion is evident in (b), this does not induce a net drift
as mentioned in the text. The data displayed above is for 
$L= 1000$.}
\end{figure}}

\section {correlations in the active-site motion}

The above description of the evolution of the interface, contained as it is
by networks of substoppers and stoppers, makes it clear that there
are strong correlations between successive points of growth or forward
motion. These correlations extend from small length scales up to 
scales of the order of the system size.
Figure 6 shows the plot of the location of  the active site for
10,000  time steps for both the tilted and  untilted cases. It can be
seen that there are jumps on all scales in the active site position.
The  figure corroborates the description of interface motion given
above. Region (1) is a typical instance of the interface
filling out a loop of size $l$. It shows that there are jumps of 
all sizes upto a
length $l$, bearing out the substopper picture. Region (2) on the other
hand marks an instance when the interface has aligned along a stopper
$p_c(m)$ and hence there are jumps of all sizes upto the 
\vbox{
\epsfysize=9cm
\epsffile{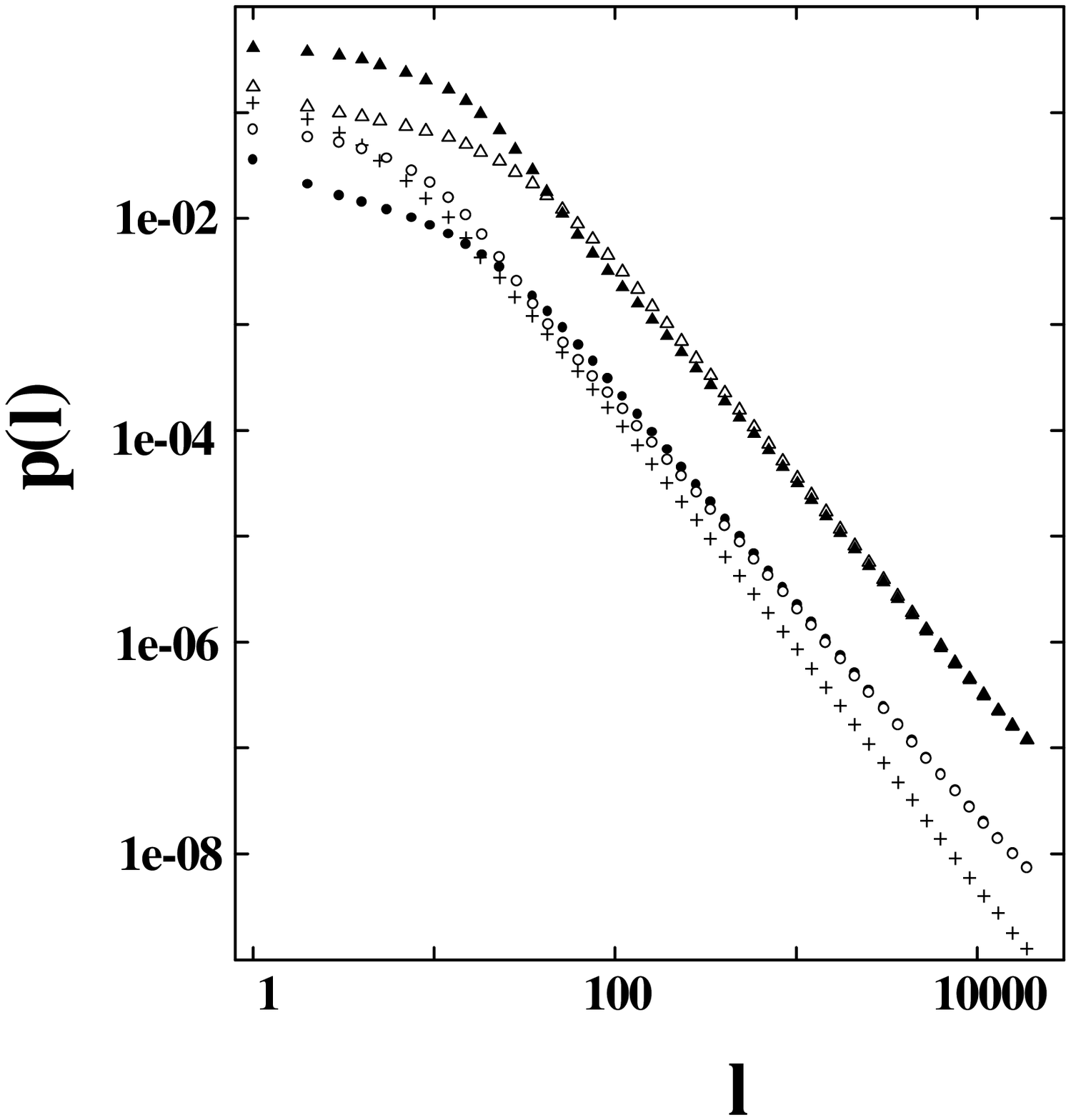}
\begin{figure}
\caption{\label{fig7}
\narrowtext
Monte Carlo results for the probability distribution 
of the jump of
the active site for three different densities, 
$\rho = 0.5$  (plus sign), 
$ \rho = 0.75$ (circles) and $\rho = 0.84375$
(triangles). If $\rho  \neq 0.5$, $p(l)$ is not a symmetric function
and $p(l)$ and $p(-l)$ (both of which are shown in the figure in the
two $ \rho \neq 0.5$ cases)
do not coincide as they do for the symmetric case. 
While $p(l) > p(-l)$ for small $l$, the curves cross so that the
situation is  reversed for large jumps. The two curves asymptotically
coincide with an asymptotic slope that differs from
that for $ \rho = 0.5$
We used  L= 65536 and averaged over $3.10^{9}$
configurations.}
\end{figure}}
system size.

More quantitatively, a measure of this long-ranged motion of the
active site  is the probability distribution $p(l)$ that two
consecutive locations of the active site 
are  a distance $l$ apart. Figure 7 
shows $p(l)$ for both the tilted and the untilted interface.  In both
cases,  
$p(l)$ decays as a power for large $l$:
$p(l) \sim |l|^{-\pi}$. 

In the
untilted  ($\rho = 1/2$) case, $p(l)$ is a purely symmetric
function  because of 
$ r \rightarrow -r $  symmetry.  We find $ \pi = 2.25 \pm 0.05$, which compares well with
earlier determined values of $\pi$ for the Sneppen model
\cite{snjn,tl1,pmb}.

In the tilted  case ($ \rho \neq 1/2$), $ p(l)$ is not a symmetric
function (Fig. 7). As can be seen from the figure, there is a
larger number of small jumps to the right, but more jumps
of large magnitude to the left. It is convenient to
separately analyse the even and odd  parts $ p_{\pm} \equiv 
(p(l) \pm p(-l))/2 $ in order to find the exponents. 
We find that the even part, $ p_{+} (l) $, decays
asymptotically as $ p_{+} (l) \sim |l|^{ - \pi_ {+}}$ with $
\pi_{+} = 2.00 \pm  0.02$. The odd part $ p_{-} (l)$ changes sign 
(as implied by the crossing of the curves in Fig. 7) and asymptotically
follows $p_{-} (l) \sim |l|^{- \pi_{-}} $ with $ \pi_{-} =2.49 \pm
0.06$. We verified that the values of
\vbox{
\epsfysize=5.3cm
\epsffile{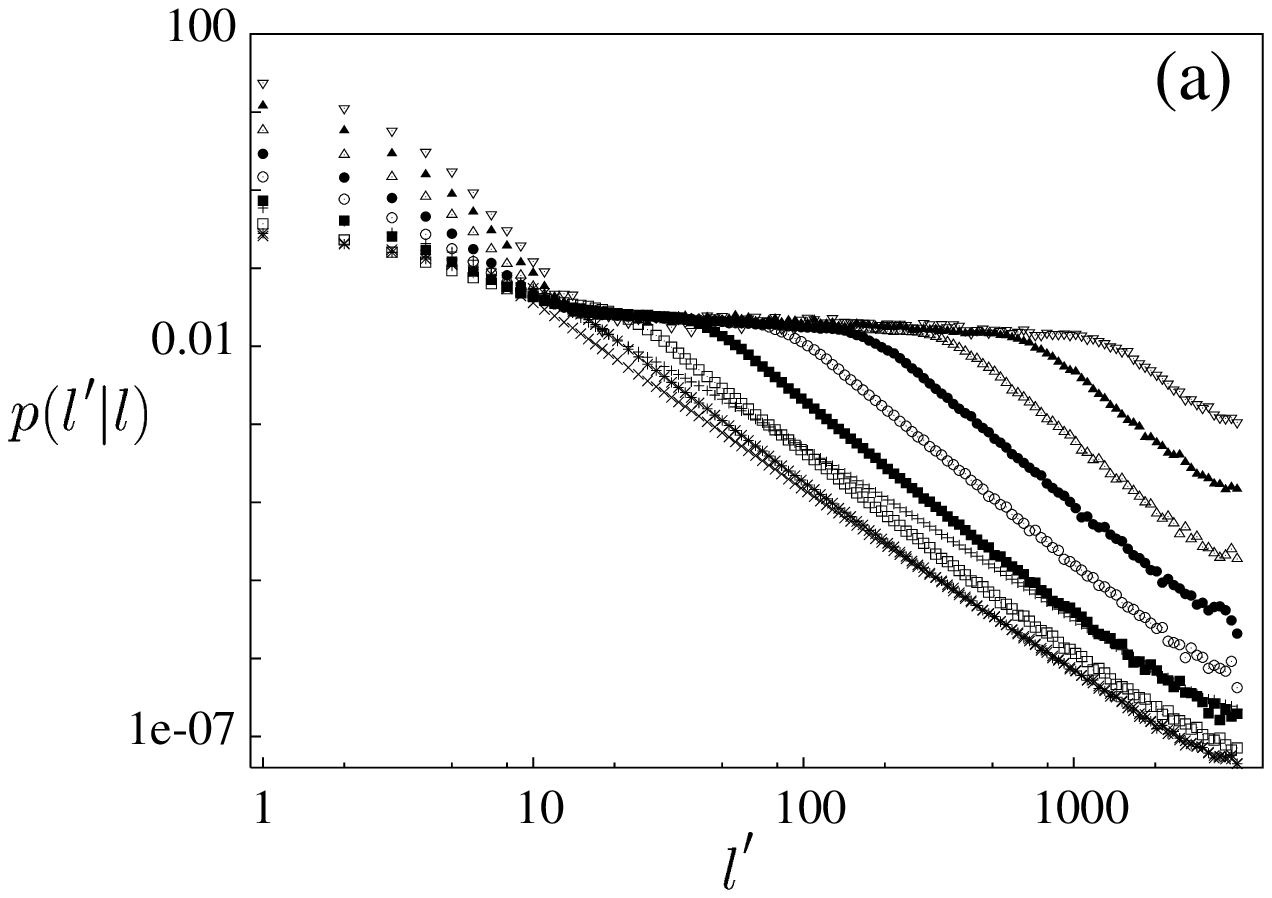}
\begin{figure}
\epsfysize=5.3cm
\epsffile{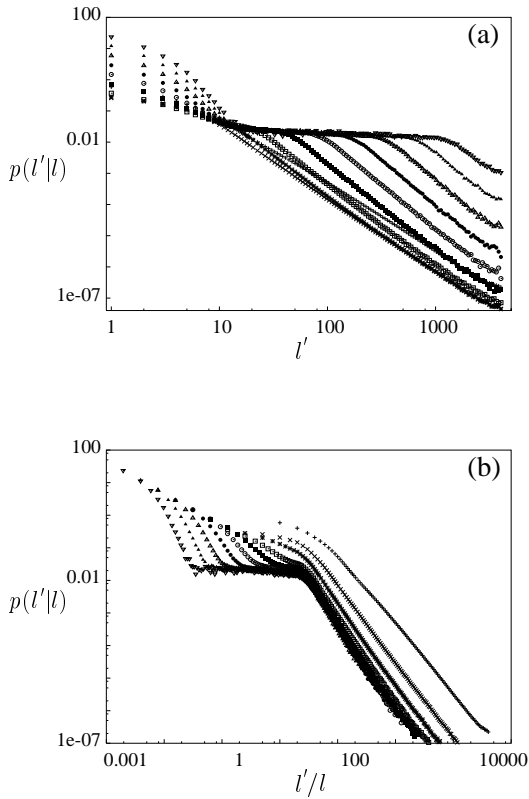}
\caption{\label{fig8}
\narrowtext
(a) The function $p(l^{\prime}|l)$ as a
function of $l^{\prime}$. 
$l$ is logarithmically binned in powers of two and is maximum for the
curve with the largest flat stretch. As can be seen in the figure, the
flat stretches of the large-$l$ curves coincide, peeling off at a
value which is $l$ dependent. 
(b) The function $ p(l^{\prime}|l)$ plotted as a function of
$l^{\prime}/l$. The curves coincide for large $l$ indicating that the
conditional probability is just a function of the ratio
$l^{\prime}/l$.}
\end{figure}}
$ \pi_{+}$ and $\pi_{-}$ are the
same for various $ \rho \neq 1/2$. 
An interesting aspect of the function $p(l)$ for the tilted interface
is that $\int lp(l) dl$ vanishes in the thermodynamic  limit.
In terms of the active-site motion, this implies that though,
for the tilted interface, the short-ranged jumps of the active site
are mostly along the tilt, there are enough long-ranged jumps in the
opposite direction to balance this. The cancellation gets better
as the thermodynamic limit is approached. The
tilt of the interface, therefore does not induce a net drift in the
active site motion.

Because of the close connection of the interface
growth problem to DP, several exponents associated with the growth and
structure of the interface are thought to be related to the DP
exponents $ \nu_{\parallel}$ and $ \nu_{\perp}$ by scaling relations
\cite{zop,tl1,pmb}. 
The exponent $\pi$ in the untilted interface as well as
the exponent $\pi_{+}$ in the tilted case have been argued to be
related to the DP exponents by the 
scaling relation  $\pi = 1 + (1+\nu_{\perp})/ {\nu_{\parallel}}$
\cite{pmb}. 
Besides this, for the tilted case, the exponents can 
also be obtained exactly \cite{maslov-zhang,tang-kardar-dhar}.

As mentioned earlier, the function  $p(l)$ does not contain
all the information about the active-site motion. This is 
due to the presence of temporal correlations in the interface
depinning model: the jump of the active site at the present instant is
strongly correlated to the jumps before. As a result, what is
needed is the full distribtion function
$p(l|l_{n-1}.....l_2|l_1)$ which is the conditional
probability distribution that a jump of length 
$l$ occurs at $t=n$ given that a jump of length
$l_{1}$ occurred at $t=1$, a jump of length $l_{2}$ occured at $t=2$ and so on. 
The probability distribution $p(l)$ is obtained by
integrating out the other variables $l_1,l_2....l_{n-1}$. In order
to assess the importance of these temporal correlations, we numerically
determined the conditional jump probability distribution
$p(l^{\prime}|l)$,  the conditional 
probabilty that a jump of length $l^{\prime}$ occurs given that at the
previous instant, the active-spot jumped a distance $l$. Figure
8a presents our numerical measurement of this function for the
EBM.  After the initial short-ranged decay, which
is a result of the local back-wetting move of 
the dynamics, the function is flat over
a considerable 
range, and then decays as a power. As evident in the figure, the
range of the flat region depends on $l$ but the value of the
probability in the flat region becomes independent of $l$ for large
$l$. In Figure 8b we show the scaling plot of $p(l^{\prime}|l)$
obtained by plotting the curves as a function of $ l^{\prime}/l$. The
function evidently approaches a scaling form for large $l$ and
$l^{\prime}$. The scaling function
is flat over a region and
decays beyond as a power $ \sim (l^{\prime}/l)^{-\pi^{\prime}}$. We
expect $\pi^{\prime} = \pi (\simeq 2.25)$ to hold as $l^{\prime}/l
\rightarrow \infty$. Although the measured value of $\pi^{\prime}$ in
the range shown is larger ($\simeq 2.9$), the bending apparent in the
lower right portion of the curve is consistent with an approach to the
value $\pi$.

Qualitatively, the behaviour of this function may be understood thus. If the
active site jumps a distance $l$
at the previous instant, one can think of the interface as 
pinned by a DP cluster with loops of average linear dimension
$\sim l$. Most of the interface would then be pinned while a portion
of it fills out a loop 
of linear dimension $\sim l$. This would imply that on an
average, any jump smaller than $l$ is equally likely.
On length scales larger than $l$ the motion is like the original
problem and hence  the jump probability
decays as a power.  This line of argument would imply that $p(l^{\prime}|l)$
should be a scaling function of $l^{\prime}/l$.  Figure 8b bears out
this expectation.  

Another equivalent way of understanding the function $p(l^{\prime}|l)$
is using the `backward-avalanche' technique introduced in ~\cite{maslov}.   
A backward-avalanche is defined as follows. If at time $s+S$ a random number 
$f_k$ is picked as the minimum, the magnitude of the backward avalanche is
$S$ if at time $s$  the random number picked was larger than $f_k$ but
for time $s+r$ for $r < S$ the value of the random number picked was
smaller. That is, one goes back in time to the 
first instant when the random number 
picked  is larger than the present one and this interval of time is
the magnitude of the backward-avalanche initiated at the present instant.
If the active site hopped a distance
\vbox{
\epsfysize=6.5cm
\epsffile{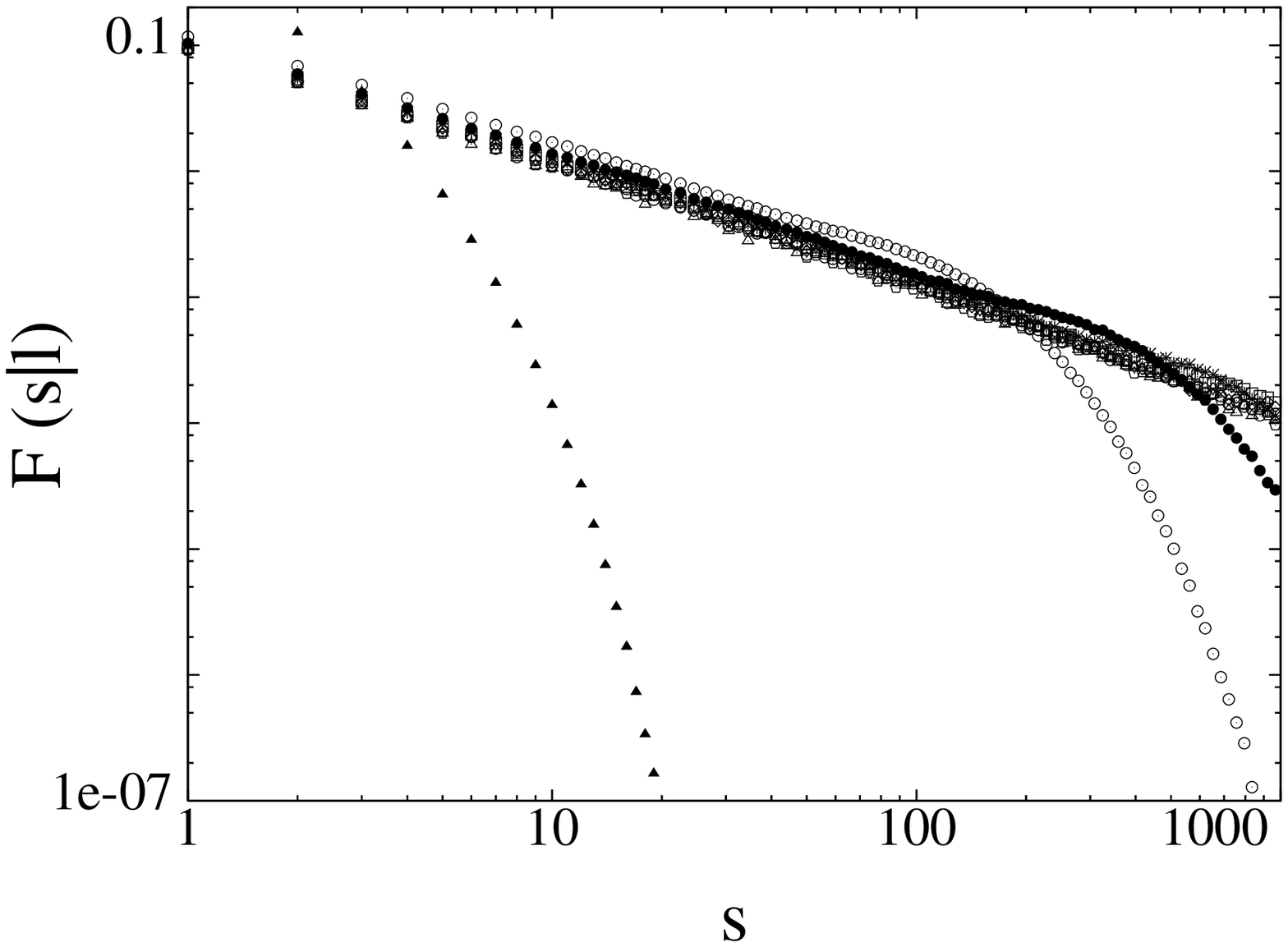}
\begin{figure}
\caption{\label{fig9}
\narrowtext
The distribution $F(s|l)$ of durations of
avalanches for jumps as defined in the text. For small $l$, $F(s|l)$
decays exponentially (filled triangles) as expected, since it is very
unlikely that a larger jump does not immediately occur. As $l$
increases, the cutoff for the decay increases (empty and filled
circles) till 
for large enough $l$, the curves coincide, decaying as a power
$\simeq s^{-1}$.}
\end{figure}}
$l$ in the previous instant, this
would imply that the size of the spatial region affected by the
backward 
avalanche  initiated at the previous instant is $\sim l$. 
 At the next
time step any of three possibilities can occur. If the
random number picked is {\it smaller} than the previous one, then 
the active site necessarily lies in the local region which has just been
affected by the move. If the random number picked is larger, then either
the random number picked belongs to the same backward-avalanche as the
previous one, or it belongs to a bigger one (which therefore
encompasses the previous one). In the former case, the active
site is equally likely to  hop anywhere in the 
region affected by the present
backward-avalanche (~\cite{pmb,maslov}) and hence the
function is flat upto a length $\sim l$. In the latter case,
active-site hops are in general larger, since the 
encompassing backward avalanche is
larger. The probability of having the active  site hop a distance
$l_1 > l$ is then related to the probability of
finding a backward avalanche of this spatial extent, and hence decays as a
power. 

Another indication of the correlation in the jumps of the active site
is the following quantity which is the analog of an avalanche, for
jumps (Fig. 9). Let the avalanche be initiated at an instant
$s$ by 
a jump of magnitude $l$ of the active spot.  The ``jump-avalanche''
lasts as  
long as consecutive jumps in the active site are less than $l$. That
is, the avalanche is of duration $S$ if at time $s$ the active site
hopped a length $l$, and for $S$ consecutive instants after that, the
hops in the active site are smaller than $l$ till at the $(S+1)$th
instant, it hopped a distance greater than $l$. 
The analog of this
quantity for random numbers was first defined for models in the
non-wetting invasion percolation regime
\cite{furubergetal,rouxetal} and later also measured for the
Sneppen model \cite{zop,tl1,pmb}.  The jump-avalanche is a strong
indication of the correlation in successive jumps of the active
site. As Fig. 9 indicates, the function falls off as a slow
power. This is an indication of the long-term memory in the system, as
the function would decay exponentially if successive jumps had not
been correlated.

\section {Pattern Formation in the Interface Depinning Model}

\subsection{Definition}

As discussed in the previous section, the motion of the active spot 
in this model is very correlated.
One can then ask whether this motion reorganizes the shape of the
interface  (or equivalently the arrangements of
particles and holes in the particle model corresponding to the EBM) in
any specific way. Below we will show that   
the correlated motion of the growth-spot in this model leads to the
formation of a pattern in height gradients 
(or equivalently the density of particles) 
of the interface.

The defining Eq. ~\ref{eq:Pattern} for height gradients can be
written in terms of densities as 
\begin{equation}
\Psi (r) = \langle n(r+R(t))\rangle - \rho .
\label{eq:pattern}  
\end{equation} 
Here $n(r)$ is the density at site $r$ and the angular brackets denote
a time average in the steady state. 

This pattern is linked to the formation of a structure which
is very different from normal space-fixed
patterns because it is referred to an origin $R(t)$ which is moving
around. In fact it
can be discerned {\it only} when the origin moves. An ordinary space
fixed average is translationally invariant and  satisfies $\langle
 n(r)\rangle  = \rho$ for any site $r$. Since the
pattern is centered around the site of activity, we will refer to it
as the ``Activity-centered pattern'' (ACP).

\subsection {Numerical results}

We studied $\Psi (r)$ by Monte
Carlo simulation in the particle-hole representation.  We studied
systems of size $L$ ranging from $2^{10}$ to $2^{16}$.  The system was
allowed to evolve through $10^6 - 10^7$ configurations before
measurements were made.  In order to speed up the algorithm to locate
the site with the minimum $f$, we used a logarithmic-bin search
procedure.   Steady state averages were computed using $\sim 10^{8}$
configurations.
 While this number of configurations was averaged over
to get an accurate estimate of the decay exponent, even an average
over about
$100$ configurations indicates the presence of a strong density
inhomogeneity clearly. 

In the untilted case, the height profile is an even function (Fig. 1a)
and $ \Psi (r)$ is an odd
function (Fig.10a)
\vbox{
\epsfysize=10.3cm
\epsffile{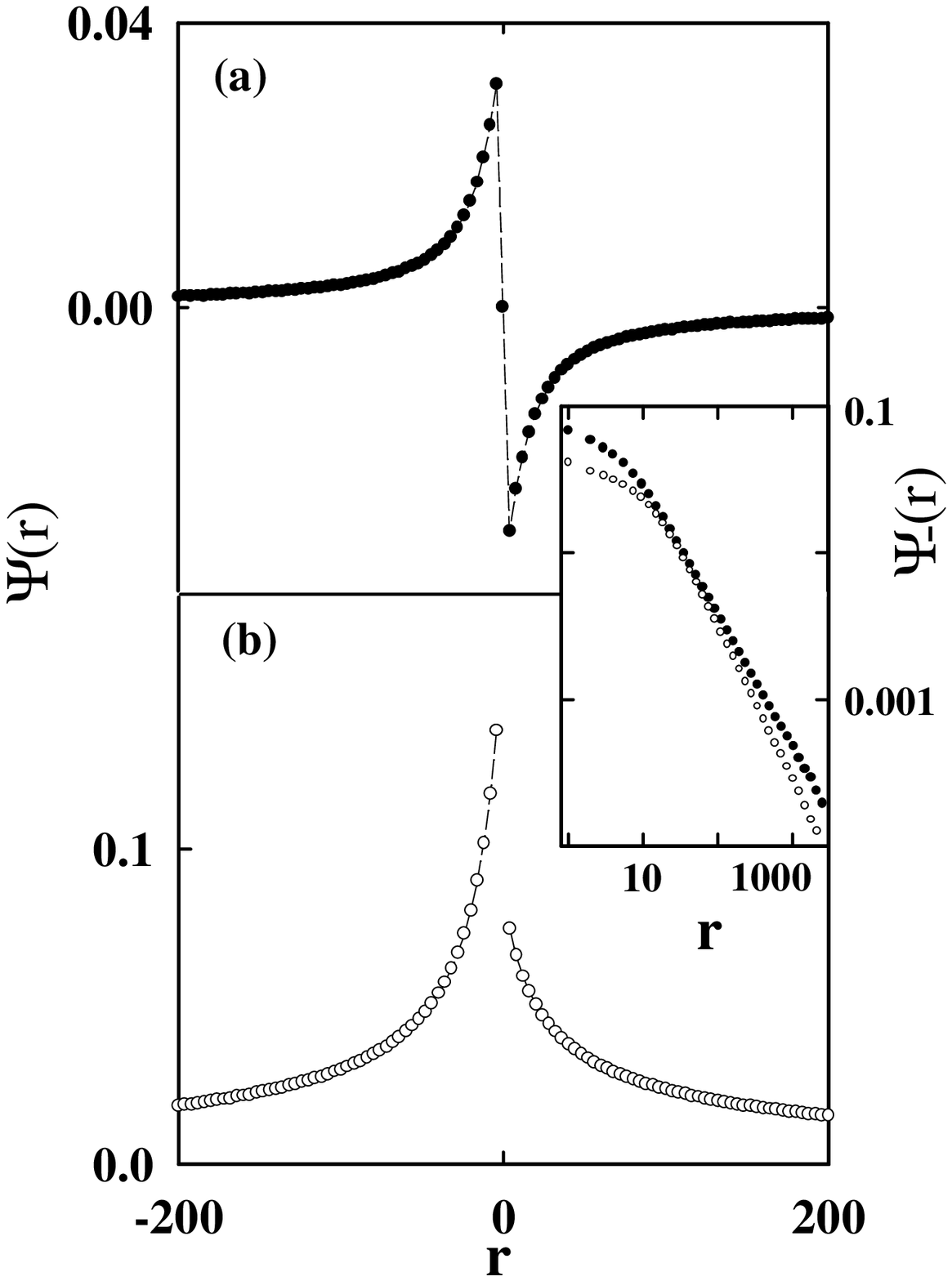}
\begin{figure}
\caption{\label{fig10}
\narrowtext
Density profiles in the (a) untilted ($\rho = 0.5$) 
and (b) tilted ($ \rho =0.75$) cases. Note that in (b), the value at
the origin is larger than $0.75$, indicating that the active site is
more than nominally likely to have a particle, because of the
bootstrap effect discussed in the text.  While (a) is an  odd function
of $r$, (b)
has no parity and the even and odd parts can be studied separately.
The inset shows the odd part of the profile in the
untilted (filled circles) and tilted (open circles) cases. Both decay
as power laws though with differing exponents.
We used $ L=16384 $ and averaged
over $ 10^{8}$ configurations.}
\end{figure}}
decaying asymptotically
as a power law $ |r| ^{-\theta}$  with
$ \theta = 0.90 \pm 0.03 $ (Fig. 10, inset).  
In the tilted case
as there is no $r \rightarrow -r$ symmetry, $\Psi(r)$ does
not have a definite parity (Fig. 10b). It is useful to separately
analyse the even and odd functions 
$\Psi_{\pm}(r) \equiv (\Psi (r) \pm \Psi (-r))/2 $. The
odd part decays as $ \Psi_{-} (r) \sim  |r|^{- \theta_{-}} $ with $
\theta_{-} = 1.04 \pm 0.05 $ (Fig. 10, inset). The even part $ \Psi_{+} (r)
\approx 
-b(L) + a ~|r|^{-\theta_{+}} $  where $ \theta_{+} = 0.46 \pm 0.05 $ and
$b(L) \rightarrow 0 $ as the lattice size $L \rightarrow  \infty$
(Fig. 11).

The density profiles of Fig.{ 10} correspond to the height
patterns shown in Figs. 1 and 2. Qualitatively, the reason for this 
time-averaged structure of the height profile can be seen as follows. 
In the untilted case, on average, the active
site is located at the peak (Fig. 1a)  where
$f_k$'s which have not been sampled earlier
are most likely to occur; such a region is thus more likely
to contain small values of $f$.
In the tilted case there is, in addition, a 
bootstrap effect at work.
If $ \rho \geq 0.5$, the active
site is more likely to contain a particle.
Given the dynamics,  
\vbox{
\epsfysize=6.2cm
\epsffile{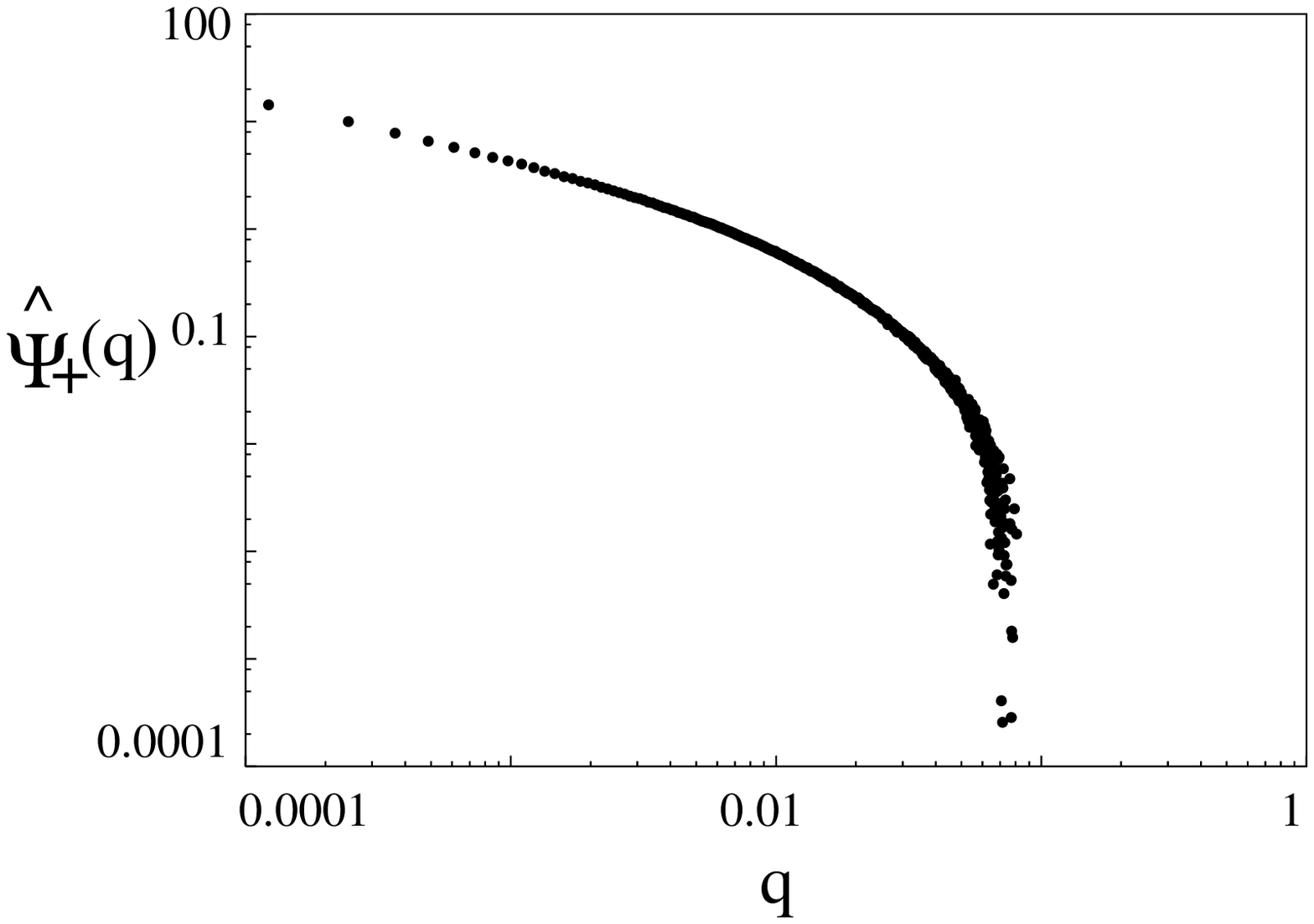}
\begin{figure}
\caption{\label{fig11}
\narrowtext
A log-log plot of the Fourier transform of the even
part of the density profile 
in the tilted ($ \rho = 0.75$) case, for the data appearing in Fig. 10. 
At low $q$ the function decays as a power $\simeq
q^{-\phi}$ with $\phi \simeq 0.54$ implying that at large $r$ , the
behaviour is a power-law with decay $r^{-(1-\phi)} $.}
\end{figure}}
regions to the left 
are more often refreshed, making
the active site likely to move in this direction
and find itself amidst a  particle cluster.  The
directionality of the active site motion is evident in Fig
{6}. This leads to densities 
higher than $\rho$ on both sides of the active site (Fig. 10b)
for $ \rho$ sufficiently different from $0.5$. That particles are
picked more often than would  be expected on the basis of the
nominal density is evident from 
the fact that the value of the density at the active-spot is larger
than $\rho$. For the density pattern in Fig. 10b, which
corresponds to $\rho = 0.75$, the value of
the density at the origin is $\Psi (0) = 0.84$.

In the 
next section we develop an approximate theoretical description for the 
exponents describing the decay of the pattern in terms of the exponent
$\pi$ which describes the 
long-ranged hops of the active site.

\section {The Integral Equation for pattern formation}

As discussed in the previous section, a pattern in height gradients
or densities is formed in the 
interface depinning model. It is  related to the motion of the
active site which  rearranges the particles and holes
in its wake. For example, if the active site  were to remain stationary
at a given site for a certain length of time, the dynamics is such
that 
it would build up a pile-up of particles 
to the left of the site and a pile-up of holes to the right {\it
i.e.} it would create a density shock around itself.
In this
section, we write down an integral equation
which provides a description of the pattern in terms
of $p(l)$, the probability that the active site hops a length  $l$ in
consecutive instants, as well as the local dynamics of interface
readjustment.

If the pattern is centered at $R(t)$ at time $t$, the dynamics 
causes two changes at the next instant:
(i) A short-ranged readjustment of the interface  changes the profile near
$R(t)$. The average density change at site $R(t)+r$ is modelled
through a density-increment function $ \Phi (r) $ which is
short-ranged.  
(ii) The active site jumps a distance $l \equiv R(t+1) - R(t) $.
Since the pattern is centered at the
active site, the result of (i) followed by (ii) is that the average
profile reproduces itself, except that it is centered at the shifted
site  $ R(t) + l $.
Both effects are incorporated into the  integral equation 
\begin{equation}
 \Psi (r) = \int_{-\infty} ^ {+\infty}  [ \Psi (r-l) + 
\Phi (r-l)] p(l) dl  \, . \label{eq:ACP}
\end{equation}
This equation can be solved using Fourier transforms.
In particular, the long-distance behaviour of the pattern $\Psi (r)$ 
is related to the decay of $p(l)$ for large $l$, resulting in a 
scaling
relation for the exponent $\theta$ in terms of the exponent $\pi$. The
details of the analysis are given in section V B below. 

However, this equation provides only an approximate description. 
To understand the nature of the approximation made, we define
a ``L\'{e}vy-flight model'', which
is similar in spirit to the models of \cite{meakinjullien,snjn}.
As in these models, there is no explicit quenched disorder, but the
effect of disorder is modelled by a long-ranged jump probability
distribution. 
For this model, we show that the integral equation (3) holds exactly.
Further, we can gain considerable insight into the mechanism of
activity-centered pattern formation by studying the effect of different
decays of the jump function $p(l)$ within 
the L\'{e}vy-flight model.

\subsection {Derivation of the Integral Equation for the
L\'{e}vy-flight model} 

The L\'{e}vy-flight model is defined as follows. Consider a
one-dimensional lattice with particles and holes.
We assume that a jump probability distribution  $p(l)$
is specified a priori: if at
$t=0$ the active site is located at a site $r$,  at the next instant,
it can lie a distance $l$ away ({\it i.e.} at $l+r$) with a
probability $p(l)$. Evidently there are no temporal
correlations in this model, since at every instant, the jump length is
chosen afresh from the distribution $p(l)$. Spatial correlations in
the active-site motion are however built in by hand since $p(l)$ is
given. Once a particle or hole is picked for update, the local rules are
assumed to be the same as in the Extremal Bond Model {\it i.e.} the
particle 
(or hole) exchanges position with the nearest hole (or particle) to
the left (or right).

We now derive an integral equation starting from
the master equation for the L\'{e}vy flight model. 
We first
define a configuration $i$ to be 
a set of integers $(\{ n _{i} (r) \}, R_{i}) ~ r =1...N$ where $N$ is
the size of the lattice and $R$ is the current position of the active
site. The variables $n(r)$ can take the values $0$ or $1$ depending on
whether the site $r$ is occupied by a particle or is empty and $R$ can
take any value between $1$ and $N$. The total number of states is, 
therefore, $ N_{T}= N \times~^{N}C_{N_p}$ where $N_p$ is the number of 
particles.
In the usual manner, we characterise the steady state by a column
vector $|P \rangle$. The entries of this column vector are the
probabilities $P_{i}$ for the configuration $i$ where $i=
1....,N_{T}$. To obtain the steady state, we need to solve the master
equation $ d |P\rangle / dt = W |P\rangle $ where $W$ is an $N_{T}
\times N_{T}$ 
matrix connecting the different states. The dynamics that connects
different states is the following:  the particle (hole) at $R$ is
exchanged with the nearest hole (particle) to the left (right) as
specified earlier. Subsequently the active site now hops from $R$ to
$R+l$ with a probability $p(l)$. The diagonal elements of the
matrix are $W_{ii} = -1 $. The off-diagonal elements are given by
$ W_{ij} = p(l)$ if 
configuration $i$ is connected to configuration $j$ by an elementary
update and a jump of the active site of length $l $
and by $ W_{ij} = 0$ otherwise . Since the
probability $p(l)$ is normalised, ($ \sum  p(l) =1 $)
the sum $ \sum_{i} W_{ij}$ for each column of the matrix adds up to
$0$. This is a requirement for any stochastic matrix.

Every configuration $i$ can go to $N $ other configurations. Each of these
differ from $i$ in the positions of the particle and hole exchanged in the
elementary update move, and also in the position of the active site. 
Similarly there are $N$ configurations which feed into any
configuration $i$. The construction of these configurations  is very
similar to that carried out in the case of the low-noise
Toom-interface and related models \cite{khop}. However, unlike in that
case, here the 
transition probabilities are not all the same. As a result the steady
state here is very different from the product-measure steady state
found in \cite{khop} and is difficult to characterise for  general
$p(l)$.

However we are able to characterise one aspect of the steady state by
defining a quantity $\Psi$ in the following manner :

\begin{equation}
\Psi (r)= \sum_{i=1}^{N_{T}} P_{i} n_{i} (r + R_{i}) - \rho \,
\label{eq:patderv} 
\end{equation}
That is, we add up the densities at a site $r$ for each configuration
that occurs in the steady state, having first shifted the origin
separately in each configuration so that the site $r$ for every
state is one which lies $r$ away from the active site. From this is
subtracted the average density $ \rho $. The above equation is
equivalent to Eq. ~\ref{eq:pattern}, with the time average in the
latter being replaced by a weighted sum over configurations  here.

From examination of the $W$ matrix, it is clear that in the case that
$p(l)$  
is a constant independent of $l$ (as in ordinary stochastic
processes), all the $P_i$'s are equal. As a result, from the
definition, the ACP vanishes. However, while this is a sufficient 
condition it is not
necessary. There are instances (see Section VIII)
when special
symmetry considerations rule out any pattern formation.
\vbox{
\epsfysize=5.8cm
\epsffile{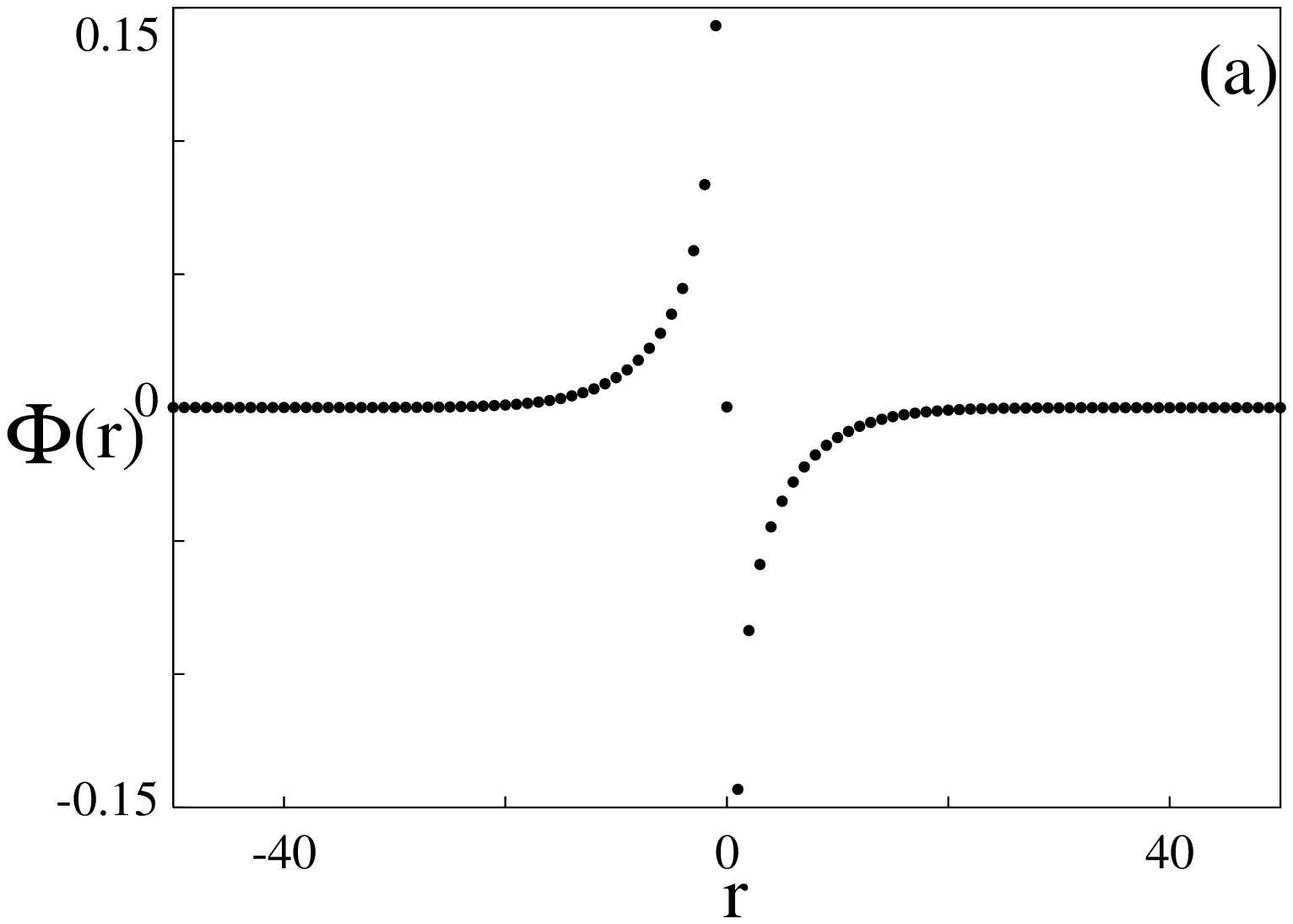}
\begin{figure}
\epsfysize=5.8cm
\epsffile{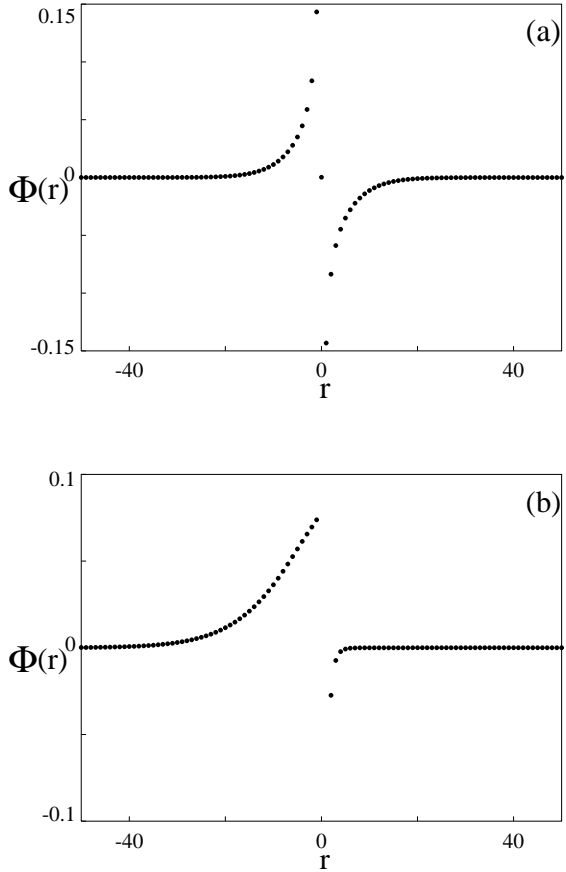}
\caption{\label{fig12}
\narrowtext
The density-increment function $\Phi (r)$ in the EBM
for (a)  
untilted 
and (b) tilted cases. Due to particle-hole symmetry, the function is 
a short-ranged purely odd function in (a) while it has no specific 
parity in (b). In both cases however, $\int \Phi(r) dr =0$ from
particle-hole conservation. We used $L= 16384$ and averaged over $16.
10^{8}$ configurations.}
\end{figure}}
However, the
presence of an ACP definitely implies the presence of a
nontrivial $p(l)$; a pattern indicates correlations in the active-site
motion.  

We will now derive Eq. ~\ref{eq:ACP}
for the L\'{e}vy-flight model.
Note that the probability $ P_i = \sum_l
p(l) \sum_{j \rightarrow i}^{\prime} P_{j}$
where the prime on the summation implies that only those configurations
$j$ are considered in which the active site is $l$ sites away from its
position in $i$. Further $j$ should transform to $i$ when the
active-site is updated. 
This equation follows from the master equation. The sum is over all
$l$'s. 

We can now substitute this in the RHS of Eq. ~\ref{eq:patderv}. 

\begin{equation}
\Psi (r)= \sum_{l} p_l \sum_i n_{i} (r + R_{i}) {\sum_{j \rightarrow
i}}^{\prime} P_{j} -\rho \, \label{eq:patderv1}
\end{equation}
The density at site $r$ in the configuration $i$ is multiplied by the
sum of the probabilities of those configurations that lead to it after
a local update; a local exchange of a particle and hole, and a
subsequent jump of the active-site of length $l$. Hence we can 
write 

\begin{equation}
n_i(r + R_{i}) = n_j(r+ R_{i} -l) + \phi_j(r+ R_{i} -l) \,
\label{eq:patderv2} 
\end{equation}
Here $\phi (r)$ is a short-ranged function (this is  related to 
the `density-increment' function appearing in Eq. ~\ref{eq:ACP}) which
depends on the local update 
rules. It is defined by Eq. \ref{eq:patderv2} and is the
difference between the two configurations $i$ and $j$ which are
related by an update. This function is the same for this model as
for the Bond model (Fig. 12) since it depends only on the local
update rules. In the EBM, it is a short-ranged function
whose range is determined by the average length of particle and hole hops. 
Later in this section we will comment on circumstances in which
this function can become long-ranged.

Substituting  Eq. ~\ref{eq:patderv2}  in Eq. ~\ref{eq:patderv1}, we
find that the right hand side can be rewritten as  

\begin{equation}
\Psi (r) = \sum_{l} p(l) \sum_{i} [n_{i} (r-l) + \phi_{i} (r-l) -
\rho] P_{i} \, \label{eq:patderv3}
\end{equation}
Using the definition of $\Psi (r)$ again, we
finally obtain the integral equation ~\ref{eq:ACP} 
for $\Psi(r)$ in terms of $p(l)$,

\begin{equation}
\Psi (r) = \sum_{l} p(l) [ \Psi (r-l) + \Phi (r-l) ] \, \label{eq:inteq}
\end{equation}
Here we have defined an averaged function $\Phi (r) = \sum_{i} P_i
\phi_i (r) $. From particle conservation, it follows that $ \int \Phi
(r) dr =0 $. For the symmetric case of half-filling, $N/2$ particles
and $N/2$ holes, the function $\Phi $ is strictly an odd function.
However this does not hold when the number of particles is not equal to
the number of holes (Fig. 12 b
). 

Thus we have been able to show that the integral equation (2) is valid
for 
the L\'{e}vy-flight model. From the nature of the model, it is clear that
while spatial correlations in the active site motion are built in by
hand, there are no temporal correlations in the length of subsequent
hops of the
active site. Therefore the integral equation is exact only in such a
case. In the next section, however, we carry over some of the
predictions of the integral equation to the EBM
and find that in some cases, it tallies quite well with 
numerical results. In Section VI, we 
discuss briefly how to generalise Eq. ~\ref{eq:inteq} to
include temporal correlations such as are present in the EBM.

\subsection {Analysis of the integral equation}

We now investigate the predictions of Eq. ~\ref{eq:ACP} for
$\Psi (r)$ in terms of a given active site hopping probability $p(l)$.
and the short-ranged readjustment function $\phi (r)$.
We solve the equation using fourier transforms.
Defining $ \hat{\Psi 
}(q) \equiv  \int_{-\infty} ^ {+\infty}  e^ {2\pi i qr} \Psi(r) dr $
{\it etc} we find
\begin{equation}
\hat{\Psi} (q) = \frac {\hat{\Phi} (q) \hat{p} (q)} {1 - \hat{p} (q)}
\, .
\label{eq:FT}
\end{equation}
\vbox{
\epsfysize=5.9cm
\epsffile{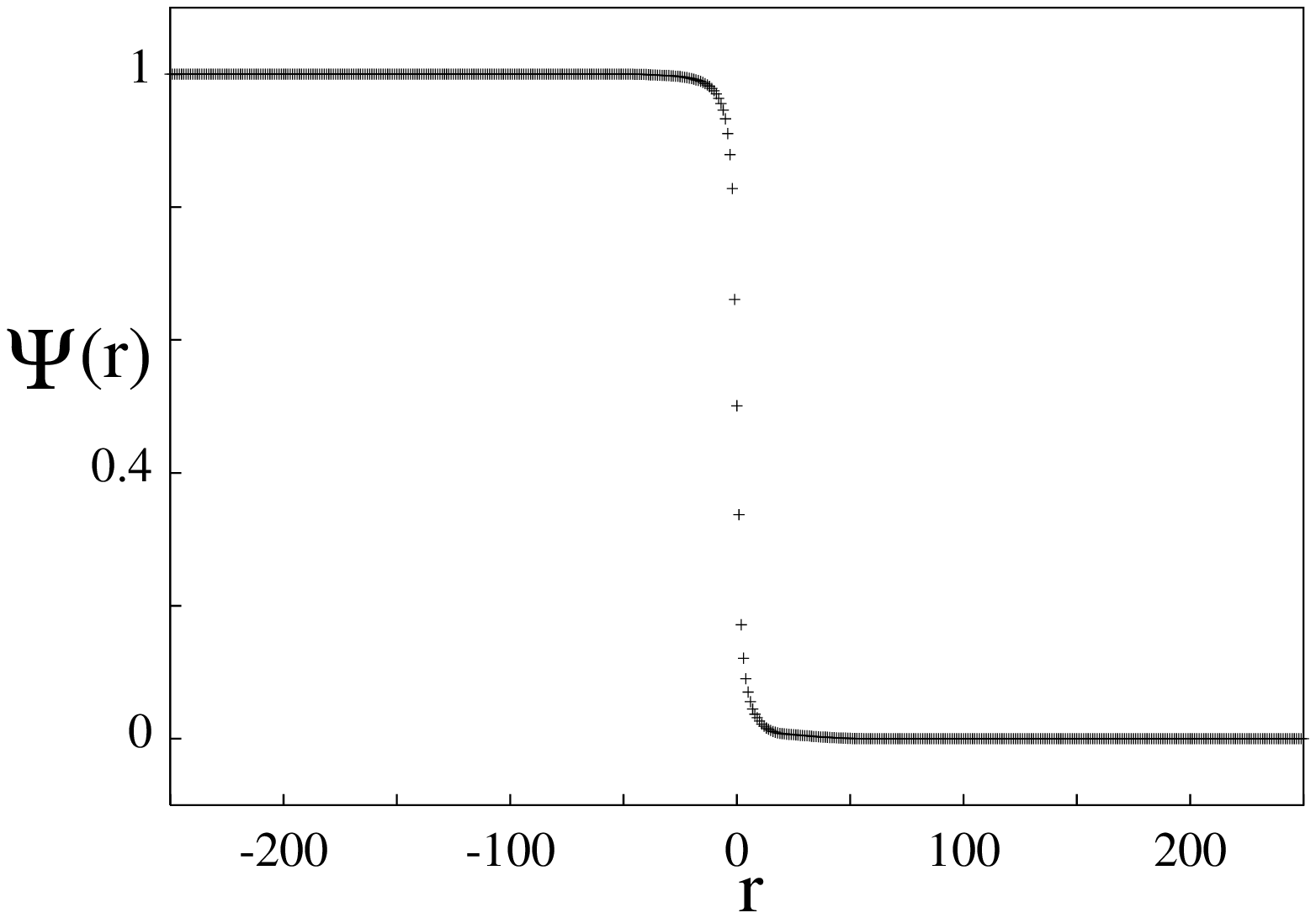}
\begin{figure}
\caption{\label{fig13}
\narrowtext
The activity-centered pattern $\Psi (r)$ in the
L\'{e}vy flight model for which $p(l)$ 
decays rapidly as $l^{-6.0}$. As evident in the figure,  
there is a near complete segregation of
particles and holes about the active site. A system of size $1024$ was
considered and $\sim 10^{5}$ configurations were averaged over.}
\end{figure}}
Since we are mainly
interested in the $q \rightarrow 0$ behaviour of  Eq.
~\ref{eq:FT}, we do not need the full functional form of
$\hat{\Phi}(q)$ but only the leading order behaviour.

Given a function $p(l)$ the integral equation predicts a corresponding
$\Psi (r)$. The large-distance behaviour of the ACP thus depends on
whether $p(l)$ is short or 
long-ranged. We consider now three different cases for the function
$p(l)$ and solve for $\hat{\Psi}(q)$ using equation ~\ref{eq:FT}. 
We substantiate the predictions of the equation by
numerically simulating the l\'{e}vy flight model. 

{\bf Case 1}: Consider first the case of an infinite-ranged $p(l)$.
The simplest 
case is when $p(l) = \frac {1}{N} $ where $N$ is the number of sites
in the lattice.  This case corresponds to usual stochastic processes.
In this case $\hat{p} (q=0)=1$ and $\hat{p} (q \ne
0)=0$. This implies that $\hat{\Psi} (q)= 0$ for $q \ne
0$ and is therefore a very short-ranged function in space.

{ \bf Case 2}: Consider now the case when $p(l)$ is a short-ranged
function. We will consider the case when it is symmetric and hence
for low $q$  $\hat p (q) \sim  q^{2}$ . Substituting this in  
Eq. ~\ref{eq:FT}
we find that $\hat {\Psi} (q) \sim  sgn ~(q)/|q|$. This implies that 
$\Psi (r) \sim sgn ~(r)$. The particles and holes separate out
completely 
and the active-site is located at the boundary between the two. This
is easy to understand if we consider the limiting case when the active
site is totally stationary. In this case, the action of the dynamics
is to move all the particles from the right of the active site 
to its left. Eventually, this leads to a total separation
of particles and holes. This picture is modified only slightly
when the active site executes a localised motion about any lattice
site, and 
hence for a short-ranged $p(l)$ (Fig. 13).

{ \bf Case 3}: We now come to the case of interest for the interface
depinning model  
{\it i.e.} when $p(l)$ decays as a slow power law. 
If $p_{+}(l) = 1/|l|^{\pi_{+}}$, Eq. \ref{eq:FT}
predicts that $\Psi (r) \sim sgn(r)|r|^{\theta_{-}}$. The exponents
$\pi $ and $ \theta_{-} $ are related by the scaling relation
$\theta_{-} + \pi_{+} =3$. This is to be compared with the 
numerical estimate $\theta_{-} + \pi_{+} = 3.15$ for the Extremal Bond
Model in the untilted case.
Consider now the case when none of the functions $p(l), ~\Psi (r)$ and
$\Phi (r)$ has a definite parity. This is relevant for the tilted
interface in the Sneppen model. 
Since $ \Phi(r) $ is short-ranged, $ \hat{\Phi} (q)
\approx i\phi_{1} q + \phi_{2} q^{2}$ as $q \rightarrow 0$.
There is
no $ \phi_{0}$ term, as the elementary step of hopping a particle or
hole conserves particle number, implying $ \int \Phi (r) dr =0$. The $ q
\rightarrow 0 $ behaviour of $\hat{p} (q)$ is determined by the
asymptotic power law decays of the even and odd parts
$p_{\pm} (l)$ as $ |l| \rightarrow
\infty$. Thus we have $ \hat{p_{+}} (q)\approx 1 - A|q| ^{ \pi_{+}
-1}$. We might have expected $ \hat{p_{-}} (q) \approx  Bq +
C ~sgn (q) ~|q|^{ \pi_{-} -1} $, but in fact the mean velocity $\int
lP(l)dl $ of the active site vanishes (as mentioned in Section III)
implying $B=0$.  
Thus the integral equation predicts that to leading order,
both $ \Psi_{+}(r)$ and $ \Psi_{-} (r) $ decay as powers $\sim $
$ |r|^{ -\theta_{\pm}}$, with 
$$
\theta_{+} + 2\pi_{+} -\pi_{-} = 3 
$$ 
and
$$
\theta_{-} + \pi_{+} = 3.
$$ 
 The prediction $\pi_{+} +
\theta_{-} =3$
compares quite well with the numerically determined values $3.04$
for $\Psi_{-}(r)$  in the
tilted case. For $\Psi_{+}(r)$, however,
the numerically
determined value of $ \theta_{+} + 2\pi_{+} -\pi_{-} (\simeq 1.97)$
deviates substantially from the predicted value $3$. The likely reason behind
the discrepancy is explained in the next section.

\subsection{Drawbacks of the Approximation}

The integral equation (2) is exact for the L\'{e}vy
flight model, but only approximate for the EBM of interface depinning.
Here we briefly run over the nature of the approximations made, and
possible directions for improvement.

One sort of approximation is the neglect of correlations between
lengths of successive jumps; as we saw in Section III, these correlations
are very marked. The extension of the integral equation to include
such correlations is discussed in Section VI B.

As noted in Case 3 of Section VB  for tilted interfaces
($\rho \ne 0.5$),  the pattern is not well represented by Eq. (2)
even qualitatively. The reason for this is that the $\rho
\ne 0.5$ pattern in the EBM is formed due to a feedback effect
which is missing in the L\'{e}vy flight model.
If $\rho>0.5$, then within EBM dynamics,
particles (or holes if $ \rho < 0.5$) are picked more often than just
$ \rho . N_{try}$ times in $N_{try}$ tries. 
This  condition is  not incorporated
in Eq.  ~\ref{eq:ACP} at all. To test how important this effect is,
we simulated the L\'{e}vy flight 
model with the further constraint that if a site chosen for growth does
not contain a particle, it  is discarded and the search is continued
till 
a site containing a particle is found. This leads to a pattern
that much more 
\vbox{
\epsfysize=5.8cm
\epsffile{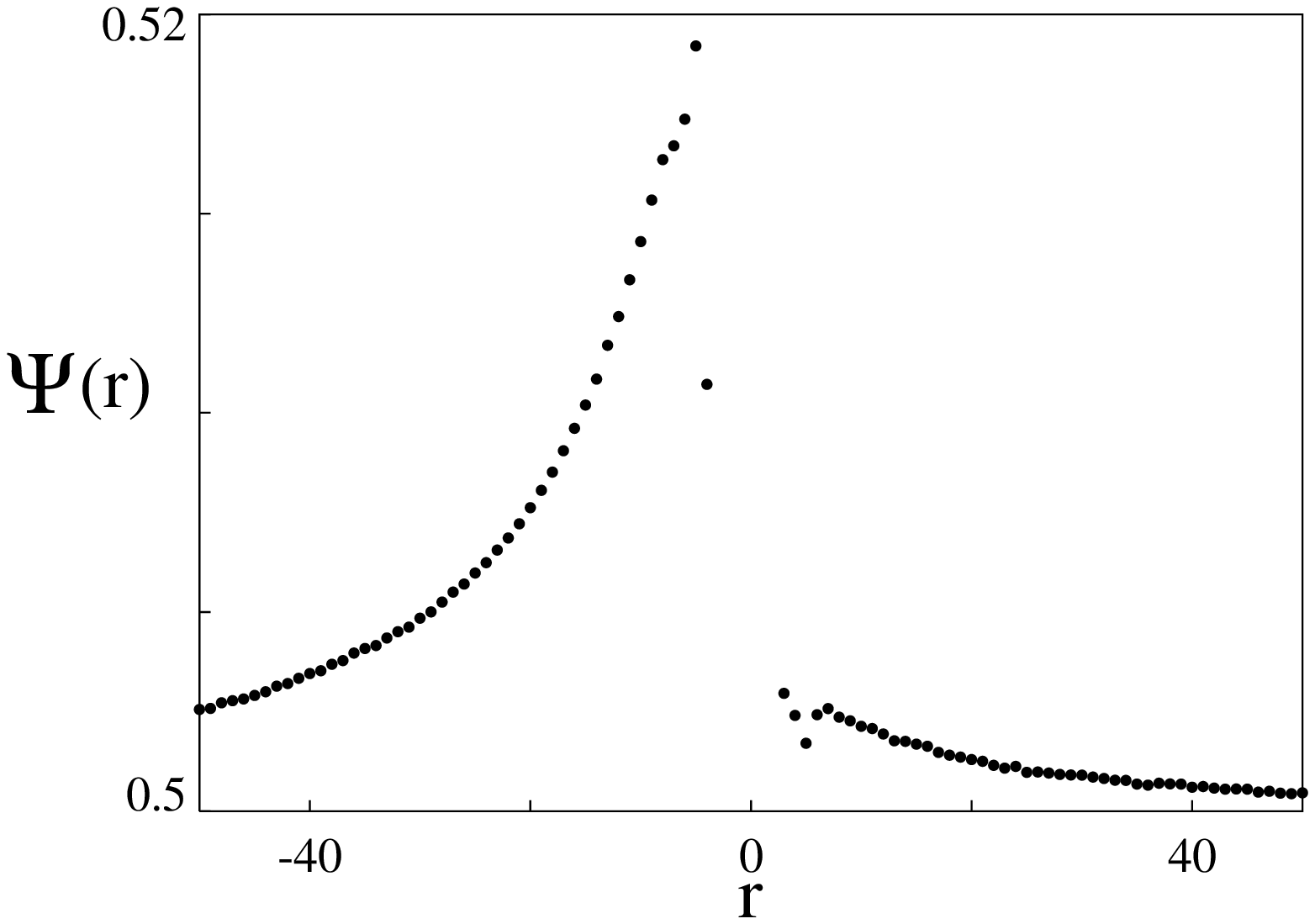}
\begin{figure}
\caption{\label{fig14}
\narrowtext
The activity-centered pattern $\Psi (r)$ for a
L\'{e}vy Flight Model with the further 
constraint that only sites containing
particles are chosen. The pattern generated with these rules 
resembles $\Psi_{+} (r)$ in the EBM more closely than if the
rules were implemented without the constraint. This fits in with 
our conjecture as to why the Integral equation does not
describe the even part of the pattern.}
\end{figure}}
resembles that in the $\rho \ne 0.5$ EBM, in that the
even part of the pattern is much more prominent than the odd 
part (see Fig. 14).     

Another point to note is that we have always taken the
density-increment function $\Phi$ to be short-ranged. Insofar as we
are interested in the EBM, this is certainly so. However, in
the L\'{e}vy-flight model, any power larger than 3, results in a
pattern like the one shown in Fig. 13, with a segregation of particles
and holes 
and not one decaying as a power as predicted by the integral
equation. The reason for this is that in this case, the mean 
squared distance $ \langle R^{2} (t)
\rangle $ covered by the active site in
time $t$ is finite. Hence for a large system, this is like a
short-ranged motion and leads to particle hole segregation. In the
process, as particle clusters build up, the density-increment function
is no longer short-ranged. 

Lastly, while in the EBM, the probability distribution of jumps
$p(l)$ arises naturally from the dynamics, in the L\'{e}vy-flight
model we put in the spatial correlation in the
active site motion by hand. In the following section, we try and
remedy this point by modifying the integral equation by writing a set
of coupled equations.

\subsection {The Coupled Equations}

A drawback of the L\'{e}vy flight model
is that the function $p(l)$ has to be supplied from the outside.
We try to rectify this by writing a set of coupled equations that
predict $\Psi (r)$ and $p (l)$ in terms of each other. 

In order to do this, we need to understand the probability 
that a certain portion of the pattern is visited by the active site. 
In the EBM, the site visited last has the largest probability of being
visited again since it is most likely to have the least random number.
The more  often a region is visited,
the more the random numbers associated with sites in that region are
refreshed and  hence the more likely it is that the minimum random
number lies in that region.
Since the dynamics creates a particle excess on the left of the active site
and a corresponding paucity on its right, a density shock is built up in 
that region.  The extent of the shock 
is directly proportional to the length of time spent by the
active spot in that region. The larger the shock, the larger the 
density difference, and hence the larger the derivative of $\Psi$.
Hence the effect of a region being 
visited a large number of times is precisely 
that $d{\Psi}/ dr$ is large.
But as explained above, the larger the number of times a site is
visited, the larger is the probability of its being visited again.
Thus we conjecture that the probability of visiting a region a distance
$l$ away from the active site is proportional to the magnitude of
$d{\Psi}/ dr$ in that region, and write the following equations:
\begin{equation}
\Psi (r) = \int dl p(l) [ \Psi (r-l) + \Phi (r-l) ] \, \label{eq:inteq1}
\end{equation}
 and
\begin{equation}
p(l)= {{\cal{N}}^{-1}} |\frac {d\Psi}{dr}|_{r=l} ~, \label{eq:coeq}
\end{equation}
where $ {\cal{N}} = \int |d{ \Psi}/dr|$  is the appropriate
normalising factor. 

Equation ~\ref{eq:inteq1} is the integral equation that we derived in
the last section. Equation ~\ref {eq:coeq} now defines $p(l)$ in terms
of the pattern $\Psi$. 
The details of the
dynamics of particle-hole exchange in this model make this a
reasonable premise as explained above.

Let us compare the results of the coupled equations with those for the
untilted interface. If $\Psi (r) \sim |r|^{-\theta}$ and $p(l) \sim
|l|^{-\pi}$, we  saw in section VB that Eq. ~\ref{eq:inteq1} implies
relation $ \pi + \theta =3$. Further, Eq.   ~\ref{eq:coeq} predicts the 
relation $ \theta +1 = \pi$. The solution is $\pi =2$ and $\theta =1$.
This compares quite well with the value of the exponents $\pi =2.25
\pm 0.05$ and $ \theta = 0.9 \pm 0.03$ found numerically for the EBM.

\section {Temporal correlations in the interface depinning model}

\subsection {Time dependence of the pattern}

The pattern, as defined earlier, is obtained by a straight time average
over all configurations, taking care to shift the origin appropriately
at every instant. It is interesting to ask how the pattern changes if 
we consider only those configurations that
occur after very large jumps of the active site (or equivalently, only those
configurations for which the minimum random number is 
\vbox{
\epsfysize=5.7cm
\epsffile{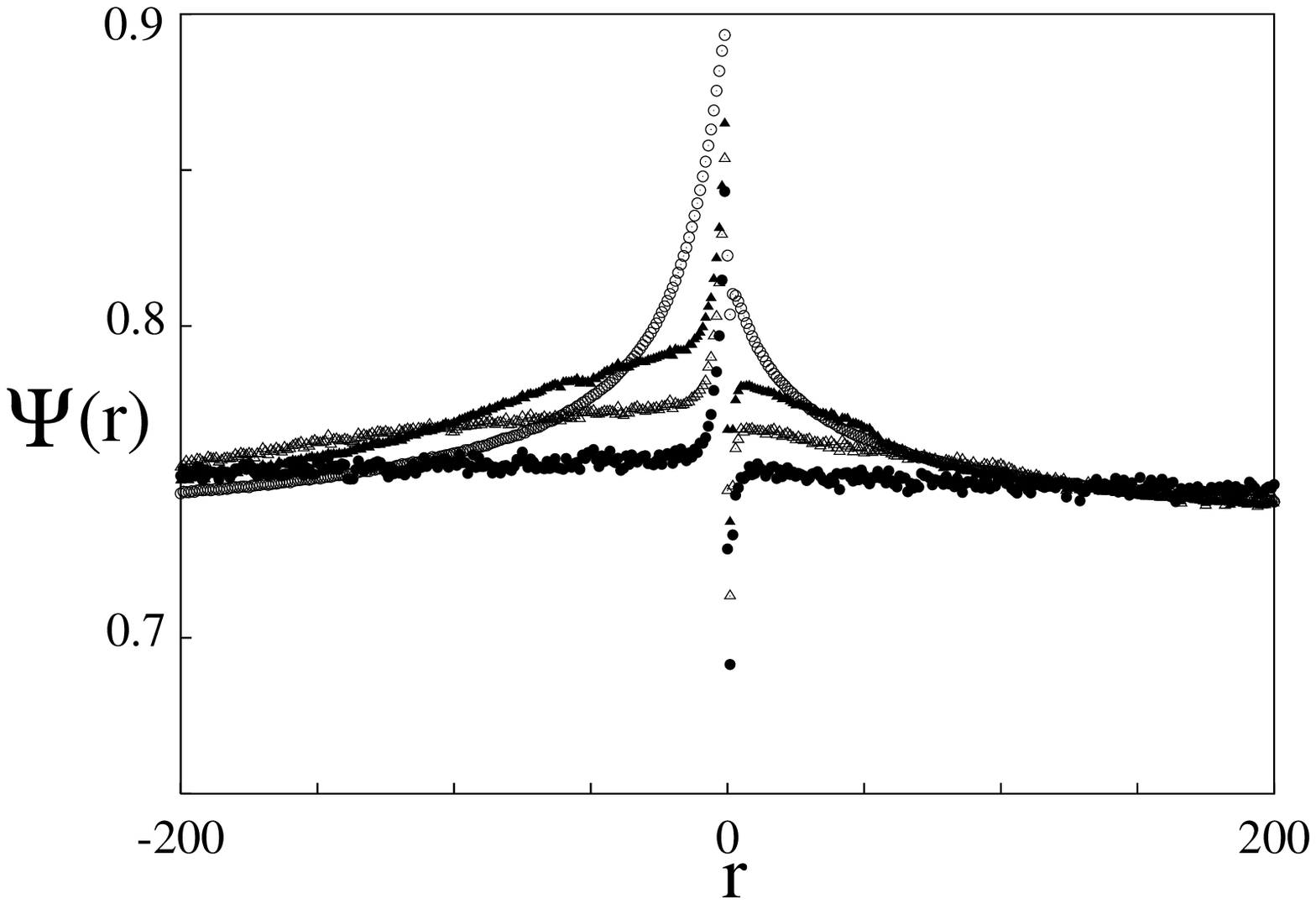}
\begin{figure}
\caption{\label{fig15}
\narrowtext
Evidence for the dependence of the ACP on the
length of the previous jump of the active site.  $\Psi (r)$ is shown
with time-averaging done over (1) all 
configurations (open circles) (2) only configurations resulting after
a jump in the active site of magnitude $50$ (filled triangles) (3)
only configurations resulting after a jump of length $100$ (open
triangles) and (4) only configurations resulting after a jump of
length $300$ (filled circles). As is evident, the pattern gets more
and more squashed till it barely exists in  (4). A system of size
$1024$ was used and $\sim 10^{6}$ configurations were averaged over.}
\end{figure}}
very close to the largest possible). 
A strong dependence of the pattern on the configurations averaged over 
is evident in Fig. 15. The pattern gets more squashed as the jumps 
in the active site leading to
the configuration get larger. This implies that the
pattern ceases to exist at stoppers and builds up again as the interface
pierces through. The actual dynamics of pattern collapse and build-up
is an interesting subject for further study.

It should be recalled that long jumps occur much more infrequently
than short jumps (Fig. 7); the probability of occurrence of a stopper
is thus very low. The time average in the definition of the pattern
(Eq. \ref{eq:Pattern}) is dominated by configurations between stoppers
(when the active site is moving within the loops of the directed
percolation network), rather than those at stoppers (when the active
site is on the backbone of the network).

\subsection {A Hierarchy of Integral Equations}

The integral equation 3 can be modified to include correlations in
time. 
To do this, we enlarge the definition of a configuration $i$ to include
the active site at the previous instant as well. Now the off-diagonal
elements of the transition matrix elements $W_{ij}= p(l^{\prime}|l)$
if configuration $j$ has resulted as a consequence of a jump of the
active site of length $l$ and is connected to configuration
$i$ by an elementary update and a jump of the active site of length
$l^{\prime}$. The function $ p(l^{\prime}|l)$ is just the conditional
jump probability already introduced in Section III.

Following the same procedure as before, we get an equation for
$\Psi_{l^{\prime}}$, the pattern resulting from an active site hop of length
$l^\prime$ :
\begin{equation}
\Psi_{l^{\prime}} (r) = \sum_{l} p(l^{\prime}|l) [ \Psi_{l}
(r-l^{\prime}) + \Phi_{l} (r-l^{\prime}) ]  ~.
\end{equation}
Keeping correlations upto one time step back gives the  pattern a
non-trivial dependence on the jump length $l^{\prime}$. The integral
equation 
~\ref{eq:ACP} gives only a trivial dependence of  $\Psi (r)$ on
$l^{\prime}$.  

This procedure can be further generalised by going back one more step
in time and keeping the location of the active site two instants
back. 

This leads to an equation of the following sort:
$$
\Psi_{l,l_1} (r) = \sum_{l_2} p(l|l_1|l_2) [ \Psi_{l_1,l_2}
(r-l) + \Phi_{l_1,l_2} (r-l) ]
$$
where $p(l|l_1|l_2)$ is the conditional probability that the active
site hops a length $l$ given that at the previous instant it hopped a
distance 
$l_1$ and in the instant before that, a length $l_2$. Similarly,
$\Psi_{l_1,l_2} (r)$ is the pattern formed when averaged over
configurations that result after two consecutive jumps of $l_1$ and
$l_2$ respectively.

Here $ \sum_{l_1} \Psi_{l,l_1} (r) = \Psi_{l} (r)$. 

Keeping the time sequence of jumps leads to an infinite hierarchy of
equations. 
\begin{eqnarray}
\Psi (r) & = & \sum_{l} \sum_{l^{\prime}} p(l^{\prime}|l) [ \Psi_{l^{\prime}}
(r-l) + \Phi_{l^{\prime}} (r-l) ] \\
\Psi_{l} (r) & = & \sum_{l_1,l_2} p(l|l_1|l_2) [ \Psi_{l_1,l_2}
(r-l) + \Phi_{l_1,l_2} (r-l) ] \\
\Psi_{l_1,l_2} & = & \cdots 
\end{eqnarray}
The integral equation ~\ref{eq:ACP} corresponds to curtailing
this hierarchy at the first step by assuming that $  \Psi_l
(r-l^\prime) = \Psi (r-l)$ and $ p(l^{\prime}|l)= p(l^{\prime}$.

\section {two-point Correlation functions in the Interface depinning
model} 
 
The activity-centred pattern $\Psi (r)$ defined in Eq. 1 is a
one-point correlation function, defined with respect to an
erratically-moving origin.  As a result,  
it has a strong effect on  the
customary, space-time averaged two-point correlation function 
$$
C(\Delta r) \equiv \{
{\langle n(r^{\prime})
n(r^{\prime}+\Delta r) \rangle} \}  - \rho ^{2} 
$$ 
Here $r^{\prime}$ is a fixed site on the lattice,
$ \langle...\rangle $ stands for a time average 
and \{...\} stands for an average over  all sites $r^{\prime}$. 

Numerical results for $ C(\Delta r)$ for a tilted interface show
that it  saturates at a value $ C^{sat}$
which decreases with 
\vbox{
\epsfysize=10.7cm
\epsffile{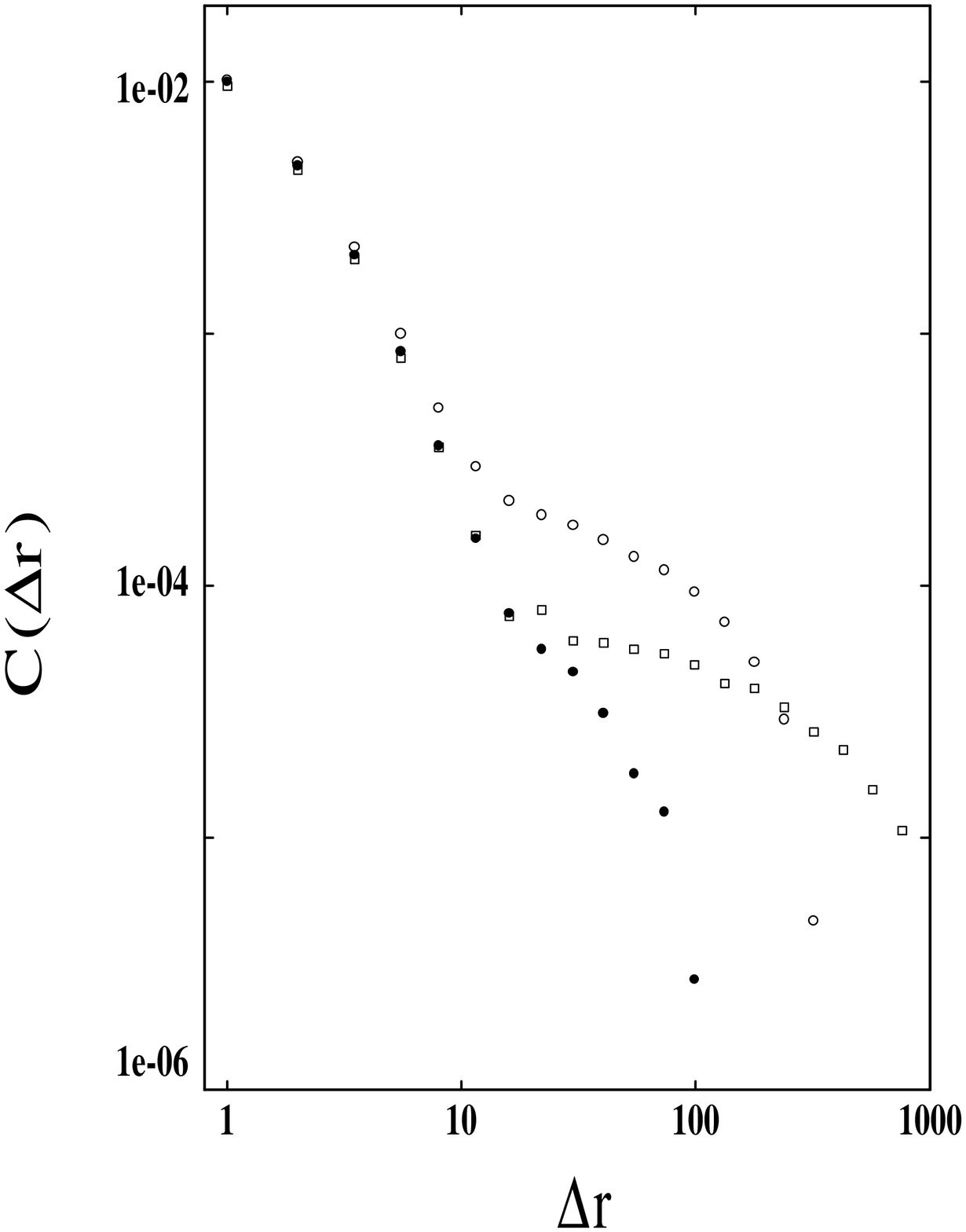}
\begin{figure}
\caption{\label{fig16}
\narrowtext
Two-point density-density correlation function for
$L=4096$ (open circles) and for $L=16384$ (squares).  The saturation
value is reduced strongly (filled circles) on subtracting the
contribution of the ACP.  This lends support to the idea that the
one-point correlation function defined by the ACP enters the
definition of the connected part of the two-point correlation
function.  The data was obtained by averaging over $10^{6}$
configurations.}
\end{figure}}
increasing size $L$ \footnote{At much larger 
values of $r$, proportional to the system size $L$, $C(r)$ changes sign,
as a result of the particle conservation sum rule  $\int C(r) dr =0$.}
(Fig 16).
This happens because
subtracting out the quantity $\rho  ^{2}$ from
the correlation function is not correct, since the presence of the ACP
causes a density inhomogeneity in the medium which cannot be accounted
for by subtracting out a constant quantity.
To account for the presence of the pattern 
we consider the correlation function 
\begin{equation}
\Gamma (r,\Delta r) = \langle
\delta n (r + R(t))\delta n( r+ \Delta r + R(t)) \rangle  \label{eq:corrln}
\end{equation}   
where $ R(t)$ is the location of the active site , $r$ is the
distance from the active site and $\delta n
(r +R(t)) \equiv n(r +R(t)) -\rho -\Psi(r)$
is the fluctuation
around the average ACP. A reasonable expectation is that the 
$\delta n$'s are independent for large separations
$\Delta r$, {\it i.e.}
$$
\Gamma ( r,\Delta r) =  \langle \delta n(r+R(t) \rangle
\langle \delta n(r+R(t)+ \Delta r)\rangle \rightarrow 0
$$ 
as $ \Delta r \rightarrow \infty$. 
Now consider averaging the function $ \Gamma ( r,\Delta r)$ over $r$.
On performing a space average over the right hand side of Eq.  ~\ref{eq:corrln},  we
note that 
$$ 
\{ \langle \delta n (r + R(t))\delta n( r+ \Delta r + R(t))
\rangle \} =  \{ {\langle \delta n(r)\delta n(r+\Delta r) \rangle} \} .
$$ 
Thus this implies that
$ \{ {\langle n(r)n(r+\Delta r) \rangle} \} -
\{ {(\rho + \Psi (r))(\rho+ \Psi(r+ \Delta r))} \}
$ approaches zero as $\Delta r \rightarrow \infty$. This
predicts the saturation value of the correlation function 
$$ 
C^{sat} = \{ (\rho+ \Psi(r))(\rho+ \Psi(r +\Delta r)) \} - \rho^{2}. 
$$ 
To test this, we subtracted this estimate of the saturation
value  from $C(\Delta r)$ 
and found that the saturation effect is in fact suppressed strongly 
(Fig. 16), supporting our interpretation.
It is possible that another slightly different definition of the
pattern would eliminate the slight shoulder which remains in Fig. 16, after
subtraction of $C^{sat}$.

Usually, if the two-point correlation function saturates as the separation
between the two points is increased, the saturation value 
is associated with a nonzero value of the space-fixed average 
$ \langle n(r^{\prime}) \rangle - \rho$. 
The unusual aspect here is that there is saturation even though
$ \langle n(r^{\prime}) \rangle = \rho$.

We studied the manner in which $C(\Delta r) - C^{sat}$ approaches zero
as $ \Delta r \rightarrow \infty$. In the tilted case, we found that
the correlation function decays exponentially, by studying a model
where the rules are the same except that only particles are
picked. 
This corresponds to a case of extreme tilt and is similar
to the case studied by \cite{maslov-zhang}. In the untilted case, $C
(\Delta r) - C^{sat}$ decays as a power law $ \sim r^{-\kappa}$ with $
\kappa \simeq 0.6$.

\section { Patterns in other models}

In the Extremal Bond Model of interface depinning, we have described
the activity-centered pattern in height gradients which forms as a 
result of correlated motion of the
active site. However, this is not the only sort of pattern
that is formed. There is pattern  formation also in the value of
the average random number $f$ at a site, as a function of the distance
from the active site.  In analogy with Eq.~\ref{eq:pattern}, 
we define this pattern in random numbers as
$$
\Psi_{f} (r) = \langle f(r+R(t))\rangle - \{ \langle f \rangle \} . 
$$
Here $f(r+R(t))$ is the random number at a distance $r$ from the
active-site and  the time average $\langle ...\rangle$ 
and space 
\vbox{
\epsfysize=4.9cm
\epsffile{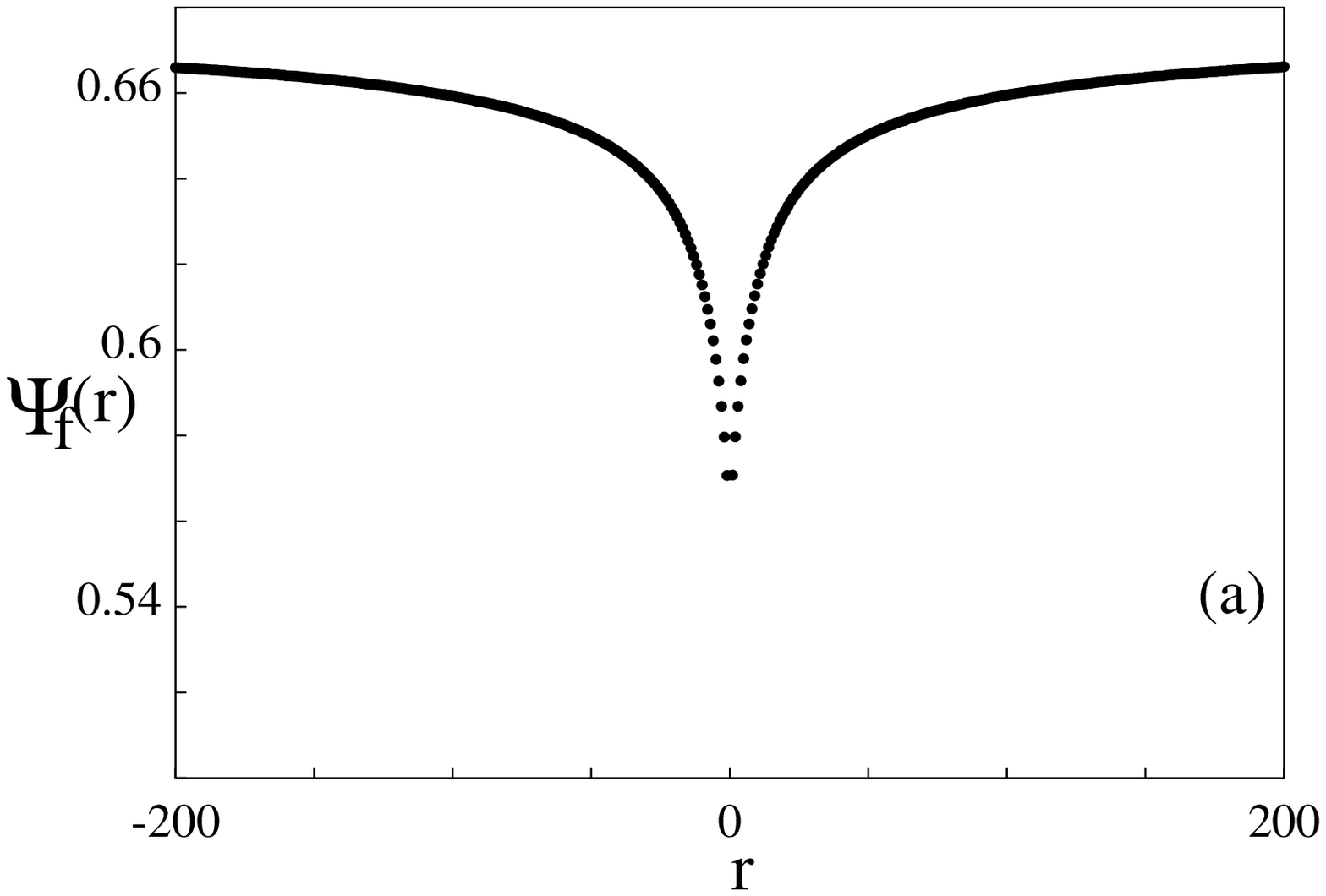}
\begin{figure}
\epsfysize=4.9cm
\epsffile{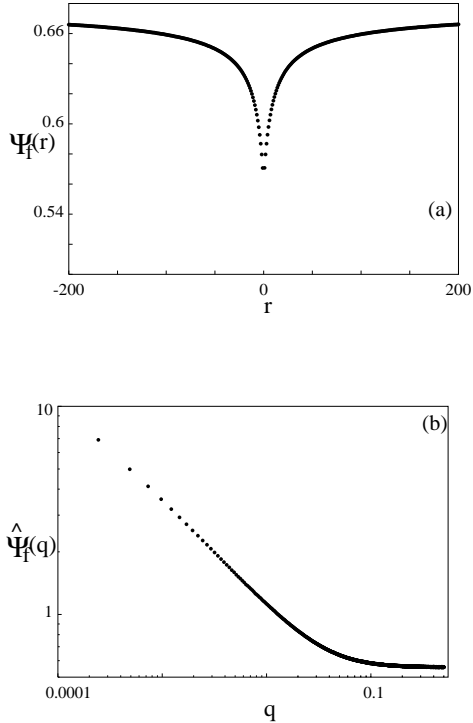}
\caption{\label{fig17}
\narrowtext
The pattern in random numbers for the EBM (the untilted
case) in (a) real space
(b) Fourier space. The Fourier transformed function appears to diverge 
as a power $q^{-\phi}$ with $\phi \simeq 0.49$ implying a power-law
decay $\sim r^{-(1-\phi)}$ at large $r$ for the 
function in real space. The above data is for a system of size
$L=8192$, averaging over $10^7$ configurations.}
\end{figure}}
average $\{ ...\}$ are  performed over configurations in
the steady state, as before.

Moreover, patterns are found in other extremal models as well. Figures
17, 18 and 19 show the $f$-patterns in the EBM, the Bak-Sneppen model
of biological evolution ~\cite{bak-sn} and the Zaitsev model of
low-temperature creep ~\cite{zaitsev} respectively.  Numerically, it
is difficult to directly extract the manner in which the patterns
shown in Figs. 17-19 approach their asymptotic values, as fits to power
law decays are very sensitive to the assumed saturation value. We
avoided this problem by studying the Fourier transforms of the
functions, as the saturation value influences only the single Fourier
mode at $q=0$. In all three cases, the Fourier transforms show
evidence of power law behaviour as $q \rightarrow 0$, implying power
law approaches of the $f$-patterns to their respective saturation
values in real space. However, we have not developed an analytical
description of the $f$-patterns in any of these models.  

Pattern formation thus seems to be generic to extremal models.
However there are instances when considerations of symmetry rule out
the formation of a pattern.  This is exemplified in the following
model, similar in spirit to that considered in ~\cite{snjnmod}. The
rules of the dynamics are the same as in the EBM, random numbers are
assigned to 
\vbox{
\epsfysize=5.4cm
\epsffile{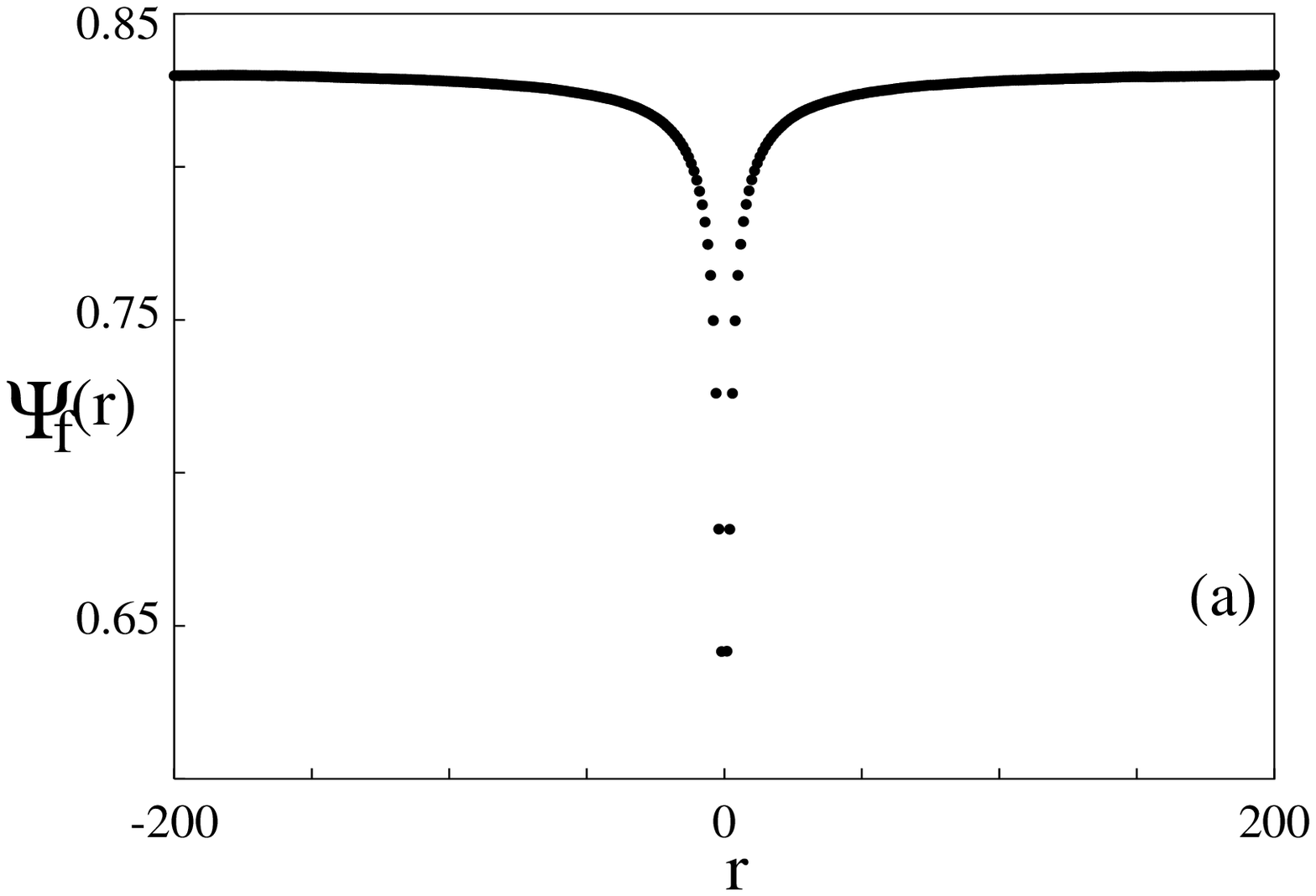}
\begin{figure}
\epsfysize=5.5cm
\epsffile{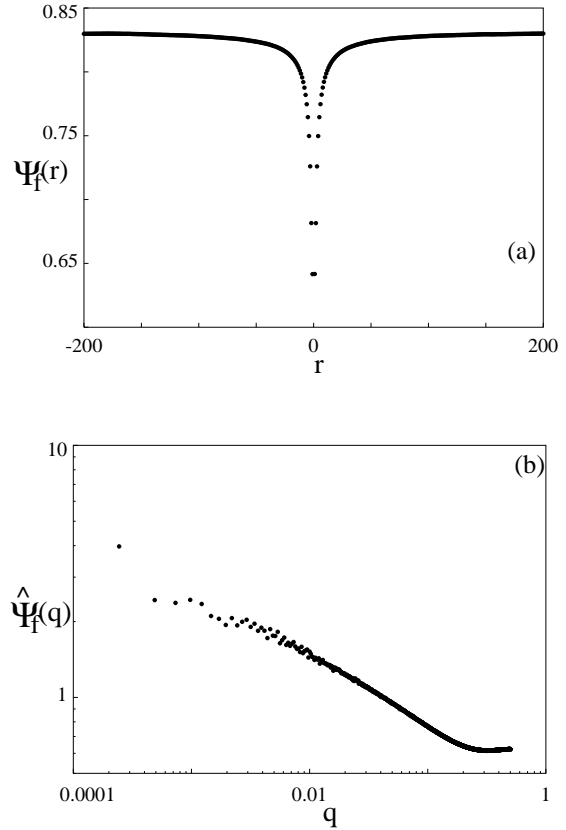}
\caption{\label{fig18}
\narrowtext
The pattern in random numbers for the Bak-Sneppen model
for evolution in (a) real space
(b) Fourier space. The Fourier transformed function appears to diverge 
as a power $q^{-\phi}$ with $\phi \simeq 0.24$ implying a 
decay exponent $(1-\phi)$ for the 
function in real space. This pattern could signify, for instance, the 
average fitness of a species as a function of the distance from the
currently mutating one. The above data is for a system of size
$L=8192$ and $  10^{7}$ configurations have been averaged over. }
\end{figure}}
every site and the minimum is picked, except that there is
no net current of particles as there was in the EBM.  If the site
picked is occupied by a particle, the particle exchanges place with
the first hole to its left, and if the site is occupied by a hole, the
hole too exchanges place with the first particle to its left.  At half
filling, there is no net current.  This model is the
most symmetric of those considered so far and there is no density
pattern formed.  However, as in the Bak-Sneppen and the Zaitsev
models, there is a nontrivial pattern in random numbers in this model.

We emphasize that the feature of the dynamics which is responsible for
activity-centered pattern formation is the existence of correlations
in the motion of the active site. Extremal models constitute just one
class in which there are such correlations. An example of another such
class is models of certain types of reaction-diffusion systems, where
the activity is quite constrained and correlated. In another physical
context, it would seem that coherent 
\vbox{
\epsfysize=5.3cm
\epsffile{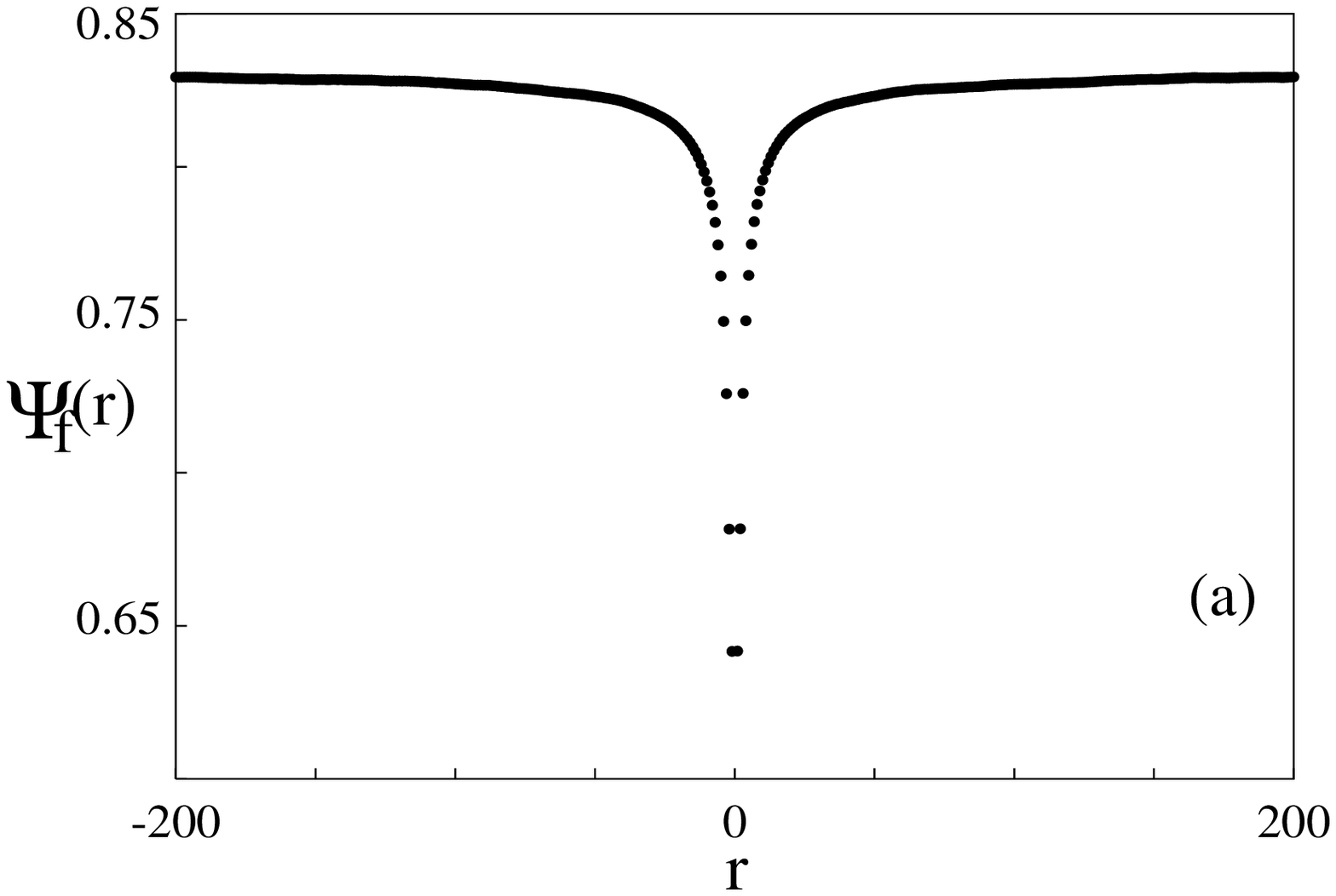}
\begin{figure}
\epsfysize=5.6cm
\epsffile{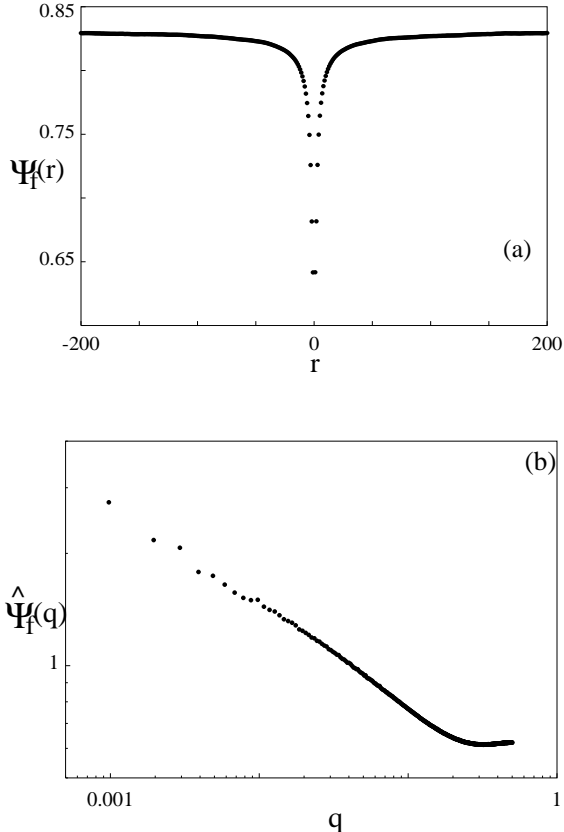}
\caption{\label{fig19}
\narrowtext
The pattern in random numbers for the Zaitsev model 
 in (a) real space
(b) Fourier space. The Fourier transformed function appears to diverge 
as a power $q^{-\phi}$ with $\phi \simeq 0.26$ implying a 
decay $(1-phi)$ for the
function in real space. The data shown is for a system of size $L=2048$,
with an average over $10^7$ configurations.}
\end{figure}}
structures which form in
turbulent flows \cite{McComb} may well be described by
activity-centered patterns. In our definition of the pattern, we take
an average over all times, keeping track of the moving structure. This
is to be contrasted with previously used methods to identify such
moving structures, based on the notion of conditional sampling,
namely, averaging only over those time zones in which the activity is
at a particular space-time location
\cite{McComb}.  Our definition takes configurations at all times into account, 
but requires a shift of the origin at every instant.

\section{Summary and Concluding Remarks}

We have introduced a variant  of the Sneppen model of interface
depinning  --- the 
Extremal Bond Model ---
and have studied the effect of the dynamics of the growth process on
the shape of the interface. Our principal result is the observation
that there is a nontrivial structure which forms in the interface,
and which moves along with the active site. A simple time average of
the height gradients, measured in a frame of reference which moves
with the active site, defines the activity-centered pattern which serves
to quantify the structure. Our numerical study shows that the pattern
has a tail which decays as a power law at large distances.

An understanding of the mechanism underlying activity-centered
pattern formation was obtained  by writing an integral equation
which relates the pattern to the probability distribution of active-site jump
lengths.
The integral equation could be derived
by writing an extended master equation for the L\'{e}vy
flight model, making it clear that the equation is exact only when there are no
temporal correlations between successive jump lengths. In the extremal
bond model, however, 
such correlations are strong. We have shown that a correct description
of the pattern, involving temporal correlations, necessitates  
keeping an infinite hierarchy of equations. Terminating this hierarchy
at the very first step results in our integral equation. It would be
of interest to understand how much one can better this description by
keeping more steps in the hierarchy. 

The activity-centered pattern is a one-point function, and so
enters the definition of two-point correlation functions.  
The physical point is that the density inhomogeneity caused by
the pattern must be taken into account, by subtracting the relevant
quantity from the density-density correlation function. If this is not
done, and the square of the 
the nominal density $\rho$ is subtracted instead, the correlation function can 
exhibit a rather
unusual sort of finite size effect. This is an interesting point
since the space-fixed time-averaged density at a site is $\rho$,
and it is only the density as defined in Eq. ~\ref{eq:pattern} 
with respect to the moving active site which is
different from the nominal density. Yet the space-fixed two-point function is
affected by this ``hidden'' pattern.

The presence of the
pattern clearly points to a non-homogeneity in the interface: the
region around the active site looks very different on average from
the region far away from it. 
For instance, we expect there to be a larger length of interface in a 
region of fixed size $x$
around the active site, than in a region opposite
it. We monitored mean squared fluctuations of the
height around the  instantaneous average,
in regions around and opposite the active site, and found a
pronounced difference (factor $ \simeq 2$, for both tilted and untilted 
interfaces
with $x = 256$,$L=4096$).
This effect is smaller at stoppers, in keeping with our finding that
the pattern itself is suppressed there.
This excess length of interface associated with
the activity-centered pattern  may provide a useful way to 
identify the active region in
experiment.     

Finally, it was pointed out that activity-centered pattern formation may occur
in a wide variety of other physical contexts, ranging from low-temperature
creep of dislocations to structures in turbulent flows. We have 
presented numerical evidence
for this sort of pattern formation in a number of other extremal models. But more generally
we expect activity-centered patterns to form whenever there are strong
correlations between successive locations of the active site.

Acknowledgements: We acknowledge helpful discussions with J. K. Bhattacharjee,
D. Dhar, G. I. Menon and M. K. Verma. We are grateful to Goutam Tripathy
for help with several of the figures.

\end{multicols}
\end{document}